\documentclass[sigconf]{acmart} 
\usepackage{hyperref}
\usepackage{float}
\usepackage{multirow}
\usepackage{color,colortbl}
\newcolumntype{L}[1]{>{\raggedright\let\newline\\\arraybackslash\hspace{0pt}}m{#1}}
\newcolumntype{R}[1]{>{\raggedleft\let\newline\\\arraybackslash\hspace{0pt}}m{#1}}
\definecolor{Gray}{gray}{0.9}
\AtBeginDocument{%
  \providecommand\BibTeX{{%
    \normalfont B\kern-0.5em{\scshape i\kern-0.25em b}\kern-0.8em\TeX}}}
\setcopyright{rightsretained}
\copyrightyear{2021}
\acmYear{2021}
\setcopyright{rightsretained}
\acmConference[CHI '21 Extended Abstracts]{CHI Conference on Human Factors in Computing Systems Extended Abstracts}{May 8--13, 2021}{Yokohama, Japan}
\acmBooktitle{CHI Conference on Human Factors in Computing Systems Extended Abstracts (CHI '21 Extended Abstracts), May 8--13, 2021, Yokohama, Japan}\acmDOI{10.1145/3411763.3443442}
\acmISBN{978-1-4503-8095-9/21/05}

\begin{document}
\title{The Challenges in Modeling Human Performance in 3D Space with Fitts' Law}
\author{Eleftherios Triantafyllidis}
\affiliation{%
  \institution{School of Informatics, The University of Edinburgh}
  \city{Edinburgh}
  \country{United Kingdom}}
\email{eleftherios.triantafyllidis@ed.ac.uk}

\author{Zhibin Li}
\affiliation{%
  \institution{School of Informatics, The University of Edinburgh}
  \city{Edinburgh}
  \country{United Kingdom}}
\email{zhibin.li@ed.ac.uk}

\begin{abstract}
With the rapid growth in virtual reality technologies, object interaction is becoming increasingly more immersive, elucidating human perception and leading to promising directions towards evaluating human performance under different settings. This spike in technological growth exponentially increased the need for a human performance metric in 3D space. Fitts' law is perhaps the most widely used human prediction model in HCI history attempting to capture human movement in lower dimensions. Despite the collective effort towards deriving an advanced extension of a 3D human performance model based on Fitts' law, a standardized metric is still missing. Moreover, most of the extensions to date assume or limit their findings to certain settings, effectively disregarding important variables that are fundamental to 3D object interaction. In this review, we investigate and analyze the most prominent extensions of Fitts' law and compare their characteristics pinpointing to potentially important aspects for deriving a higher-dimensional performance model. Lastly, we mention the complexities, frontiers as well as potential challenges that may lay ahead.
\end{abstract}



\maketitle

\section{Introduction}
Fitts' law is the most widely used human performance model in HCI history \cite{GILLAN1992291, 10.1207/s15327051hci0701_3,10.1145/1753846.1753867,doi:10.1080/00140139108967324,argelaguet2013survey,doi:10.1080/00222895.1973.10734962,so2000effects}. With the growth of networking and mixed reality technologies, the necessity of assessing human performance with a higher dimensional model has increased significantly \cite{9076603, 10.1145/1753846.1753867}. Fitts' proposed his original model, known as simply Fitts' model in 1954 \cite{fitts1954information}. Its applicability has been well demonstrated in 2D tasks \cite{fitts1954information, 10.1207/s15327051hci0701_3, doi:10.1080/00140139508925153}, motivating multiple other researchers with their own 2D extensions, most notably that of Hoffmann's, Welford's and Shannon's \cite{10.1145/302979.302989, welford1968fundamentals, 10.1207/s15327051hci0701_3}. More recently, the law itself has also been tested in 3D space \cite{doi:10.1177/0018720810366560, 8998368} and providing a robust ground for various 3D extensions \cite{murata2001extending, 10.1145/3290605.3300437, CHA2013350}, with lesser but still impressive predictive powers. 

It would thus come naturally that somewhere out there, there would exist an extension of Fitts' law to model the entirety of the 3D domain. However, to the best of our knowledge, this is not the case. The most popular methods to date have provided an invaluable insight as to which variables are integral when spatially describing 3D space \cite{10.1145/302979.302989, welford1968fundamentals, 10.1207/s15327051hci0701_3, murata2001extending, CHA2013350, doi:10.1177/0018720810366560, 10.1145/3290605.3300437, 8998368}. Consequently, these methods were tested under very specific settings and limited to certain spatial arrangements, thus progress towards a true 3D performance metric is still scattered not allowing for inter-study comparisons. More specifically, a few models accounted for varying gains of spatial arrangements including directions and inclinations \cite{8797975, 10.1145/3290605.3300437, murata2001extending, CHA2013350}, of which only the latter two formulated an extension. While Fitts' model was originally intended for translational tasks only, other studies demonstrated that it can adequately model 2D rotational tasks as well \cite{412031, doi:10.1080/00140136708930874, doi:10.1080/14640748308402133}, though not exhaustively tested in the 3D domain. Furthermore, combined translational and rotational movements are severely limited and only accounted by two studies \cite{doi:10.1177/0018720810366560, 8998368}, of which movements were limited only across one line, effectively disregarding spatial arrangements. Moreover, depth-related distances in 3D displays were only accounted for and added to a model extension in one study \cite{10.1145/3290605.3300437}. Last but not least, with the exception of Hoffmann's \cite{doi:10.1080/00140139508925153} all aforementioned studies only considered the effective target size, effectively ignoring the probe size i.e. the object size used to point to the target location. All of these studies reported significant influences to their models, yet with the absence of a standardized metric, inter-study validations remain particularly challenging.

We can thus infer that there are numerous spatial variables and multiple complexities arising when trying to extend Fitts original model to 3D space. With this review, we hope to investigate and pinpoint towards important factors one has to consider if attempting to propose a model in 3D space, taking into account varying gains of translational and rotational complexities including spatial arrangements.

\section{Background and Related Work}
In this section, we investigate and analyze the most prominent and widely used extensions based on Fitts law. Moreover, we compare all formulations and their applicability in increasing spatial complexities as to pinpoint towards which directions researchers should direct focus to if attempting to derive a full 3D performance model. In \autoref{table:ModelComparisons} we summarize all models and extensions as an overall visual overview of their applicability towards higher dimensions.

\subsection{Original Law}
Fitts' original formulation \cite{fitts1954information, fitts1964information} predicts the \textit{movement time} (MT) based on an index of difficulty (ID) and is formulated as follows:
\begin{equation}
    \label{eqn:fitts_law_original}
        \begin{split}
            MT = a + b \cdot ID, \\ ID = log_2 \left ( \frac{2A}{W} \right )
        \end{split}
\end{equation}
It can be thought of as the time to reach/point or click to a target location, given the target's distance (A) from the origin of the cursor/hand or object, as a ratio of the target's width (W). The logarithmic term \(ID\), represents the \textit{index of difficulty} measured in bits per second [bit/s] while \(MT\) is measured in seconds. Constants \(a\) and \(b\) represent the intercept and slope respectively and are derived via regression. In the following sections, we will present the most widely used extensions of Fitts' model, also indicating the ID for each equation to make it clear; as with most work, it is not clearly stated which part represents the actual ID. 

\subsection{Extensions in 2D Space}
Including Fitts' formulation, there are numerous variants based on his extensions, including: 
\begin{equation}
    \label{eqn:fitts_law_original_MT}
            MT = a + b \cdot \underbrace{log_2 \left ( \frac{2A}{W} \right )}_{\textit{ID}}
\end{equation}
\begin{equation}
    \label{eqn:hoffmann}
            MT = a + b \cdot \underbrace{log_2 \left ( \frac{2A}{W+F} \right )}_{\textit{ID}}
\end{equation}
\begin{equation}
    \label{eqn:welford}
            MT = a + b \cdot \underbrace{log_2 \left ( \frac{A}{W} + 0.5 \right )}_{\textit{ID}}
\end{equation}
\begin{equation}
    \label{eqn:shannon}
            MT = a + b \cdot \underbrace{log_2 \left ( \frac{A}{W} + 1 \right )}_{\textit{ID}}
\end{equation}
\autoref{eqn:fitts_law_original_MT} represents Fitts' original formulation. \autoref{eqn:hoffmann} is an extension of Fitts' by Hoffmann  \cite{doi:10.1080/00140139508925153}. In his formulation, Hoffmann kept Fitts' law mostly intact, with the exception of adding the variable \(F\), representing the index finger pad size of each participant. This stemmed from a series of experiments he conducted mainly composed of discrete tapping tasks using the participants' pad finger size of their index as pointing probes. His formulation opened new and interesting paths. Indirectly, it motivated future work in including the object size when concerned with manipulating objects in virtual environments \cite{10.1145/302979.302989}. \autoref{eqn:welford} presents Welford's extension \cite{welford1968fundamentals}, removing the multiplication by 2 for the target separation i.e. distance (A) but adding +0.5 in his formulation. Similarly, in \autoref{eqn:shannon} Scott MacKenzie introduced the Shannon's formulation  \cite{10.1207/s15327051hci0701_3}, similarly to Welford's but instead of 0.5 adding a plus +1 term. The difference between the latter two lays in the added terms. Welford, contrary to MacKenzie, argued that the reason for adding the +0.5 term was to account for the distance from the centre of the target to its edge.  Based on the Shannon Formulation \cite{10.1207/s15327051hci0701_3}, another model extension, named FFitts Law \cite{10.1145/2470654.2466180}, was suggested and formulated as:
\begin{equation}
            MT = a + b \cdot \underbrace{log_2 \left ( \frac{A}{\sqrt{2 \pi e \left (\sigma^2 - \sigma_{\alpha}^2 \right )}} + 1 \right )}_{\textit{ID}}
\end{equation}
replacing the denominator, effective target width (W), with a double Gaussian distribution in which $\sigma$ represents the standard deviation of the touch points and $\sigma_{\alpha}$ the precision of the input finger. This approach showed a promising accuracy in both 1D and 2D target acquisitions tasks.

Nonetheless, the most widely used 2D extension of Fitts law still remains that of MacKenzie's i.e. the Shannon formulation and has been demonstrated to do very well for tasks entailing purely translational  \cite{doi:10.1080/00140139.2011.614356, 10.1145/1357054.1357308, doi:10.1177/154193120905301216, 10.1145/985692.985749, CHA2013350, doi:10.1177/0018720810366560, 7546014} or rotational settings \cite{meyer1988optimality, doi:10.1177/0018720810366560}.

\subsection{Extensions in 3D Space}
Fitts' formulation has also been applied in the 3D domain but has shown not to represent 3D movements accurately \cite{10.1145/2395131.2395135, 10.1145/2858036.2858244, 8797975, doi:10.1177/0018720810366560, doi:10.1177/1071181312561207}. Consequently, certain extensions were needed to account for this limitation \cite{10.1007/978-3-642-39330-3_38, 5444713, 10.1007/978-3-642-34182-3_2, CHA2013350, murata2001extending, 10.1145/3290605.3300437}. The first being directions. Murata and Iwase \cite{murata2001extending}, were the first to introduce directional angles in their study, in their case phrased as azimuth angles under the spherical coordinate system. A total of eight different levels of directional angles were investigated ranging from 0$^{\circ}$ to 315$^{\circ}$ with a 45$^{\circ}$ increment. Their findings showed that these angles had a sinusoidal relationship with movement time. More specifically, they found that upper (90$^{\circ}$) and lower movements (270$^{\circ}$), were significantly more difficult and by extent increasing MT than left (180$^{\circ}$) or right movements (0$^{\circ}$). Cha and Myung \cite{CHA2013350}, inspired by Murata and Iwase's model \cite{murata2001extending}, proposed an additional term, inclination. In their experiment, they introduced both varying gains of directional as well as inclination angles for pointing tasks. They confirmed Murata and Iwase's work in that directional angles do indeed appear to have a sinusoidal relationship with MT and moreover, inclination angles appeared to have an almost linear relationship with MT. Neither of those models however investigated their findings within an experimental setting of 3D displays, let alone VRHMDs. This was later accounted for by Machuca and Stuerzlinger \cite{10.1145/3290605.3300437}. The latter investigated the stereo deficiencies in virtual hand pointing with the use of 3D displays. Foremost, they confirmed that left to right movements were significantly easier than movements away from or towards the user. However, it is important to mention that both Murata \& Iwase \cite{murata2001extending}, as well as Cha \& Myung \cite{CHA2013350}, studied this discrepancy with the directional azimuth angles being perpendicular to the view direction of the participant, i.e. a frontal circle in front of them. Whereas in Machuca and Stuerzlinger's, the directional angles were placed around the participant with 90$^{\circ}$ and 270$^{\circ}$ representing the front and backward whereas 0$^{\circ}$ and 180$^{\circ}$ degrees represented left and right movements respectively. The most important however finding was that depth changes i.e. the distance of the user's eyes to the screen, linearly affected MT, with higher depth values presenting higher difficulties and by extent higher timings. The aforementioned models are shown below.
\begin{equation}
\label{eqn:murata_inclination}
    MT = a + b \cdot \underbrace{\left ( log_2 \left ( \frac{D}{W} + 1 \right ) + c \cdot \sin{\theta} \right )}_{\text{ID}}
\end{equation}
\begin{equation}
\label{eqn:cha_myung_incline_azimuth}
    MT = a + b \cdot \theta_{1} + c \cdot \sin{\theta_{2}} + d \cdot \underbrace{log_2 \left ( \frac{2D}{W + F} \right )}_{\text{ID}}
\end{equation}
\begin{equation}
\label{eqn:Machuca_stereo_deficiencies}
    MT = a + b \cdot  \underbrace{log_2 \left (\frac{A}{W} + 1\right )}_{\text{ID}} + c \cdot CTD
\end{equation}
Murata and Iwase's \cite{murata2001extending} directional model is shown in \autoref{eqn:murata_inclination}. The variable $\sin{\theta}$ represents the sinusoidal directional / azimuth angle, controlled by a constant c, determined through regression. Note that in Murata and Iwase's model, contrary to other extensions, the ID not only encompasses the logarithmic but also the sinusoidal term with the azimuth angle added to it. Cha and Myung's model is shown in \autoref{eqn:cha_myung_incline_azimuth}. Contrary to Murata and Iwase's model which is based on Shannon's, Cha and Myung's \cite{CHA2013350} is based on Hoffmann's taking into account the finger pad size of the participants acting as the pointing probe (F). In addition to the directional angle ($\sin{\theta_{2}}$), they introduced $\theta_{1}$ which represents inclinations sharing a linear relationship with MT. Constants a,b,c and d are again determined through regression. Finally, Machuca and Stuerzlinger's model \cite{10.1145/3290605.3300437} is shown in \autoref{eqn:Machuca_stereo_deficiencies}, which is based on Shannon's \cite{10.1207/s15327051hci0701_3} shown in \autoref{eqn:shannon} with the addition of CTD representing the Change in Target Depth (measured in centimeters), controlled by a constant c determined through regression. None of the aforementioned models in this section, however, included combined translational or rotational variations.

\subsection{Translation or Rotation?}
To this point, all the formulas reported, either extended the original Fitts' law from 2D to 3D space, but were solely limited to translation. However, during object interaction, be it pointing or manipulation, rotation is a fundamental part. When performing a task that requires some kind of spatial accuracy, we humans usually attempt to match the rotation of the object so it satisfies certain spatial criteria \cite{9076603}.

Stoelen and Akin \cite{doi:10.1177/0018720810366560}, were the first to combine both translational and rotational movements in one experiment. Motivated by MacKenzie's Shannon formulation \cite{10.1207/s15327051hci0701_3} shown in \autoref{eqn:shannon}, Stoelen and Akin proposed that simply adding the indices of translation and rotation would yield adequate results in modeling combined movements. To adjust the formula of translational movements of Shannon's, Stoelen and Akin replaced the otherwise target distance (A) as the numerator, with the respective rotational distance ($\alpha$) and the denominator (W) indicating the target width, with the rotational tolerance ($\omega$). Their formulation is shown in \autoref{eqn:stoelen_combined}:

\begin{equation}
\label{eqn:stoelen_combined}
\begin{split}
    MT_{combined} = a + b \cdot \left ( ID_{translation} + ID_{rotation} \right) \\
    ID_{translation} = log_2 \left ( \frac{A}{W} + 1 \right) \\
    ID_{rotation} = log_2 \left ( \frac{\alpha}{\omega} + 1 \right)
\end{split}
\end{equation}

As such, the total combined movement time ($MT_{combined}$) it takes to point to a target is dependent upon the index of difficulty for translation ($ID_{translation}$) and rotation ($ID_{rotation}$), sharing the same "weight" and linearly correlated to MT. Stoelen and Akin, however, limited their findings without any spatial arrangements since pointing to the target was performed only across one line and furthermore performed in 2D space without the use of 3D displays \cite{doi:10.1177/0018720810366560}. The latter was accounted for by Kulik et al. \cite{8998368}, whereby 3D displays were used to model combined transitional and rotational movements. They found a surprisingly adequate linear fit of $R^2=0.78$ when combining both translation and rotation, defined in their case as "3D Docking" \cite{8998368}. They confirmed the findings of Stoelen and Akin yet movements were again limited along one line only. The question however remains, does a rotation in 3D space really share a linear relationship with MT as with translation? With the exception of the two studies mentioned above \cite{doi:10.1177/0018720810366560, 8998368}, this is not sufficiently investigated and perhaps rotation could share a polynomial or exponential relationship with MT. It would thus be invaluable to further confirm their findings.

\subsection{Pointing or Object Manipulation?}
In the larger context of object interaction, there are in general two major categories, pointing and manipulation. Until now, we investigated purely pointing tasks, that is without the presence of any physical interactions such as gravity or contact points. One could argue that covering pointing tasks under a unified 3D performance model would suffice. However, use cases encompassing teleoperation and in general simulation training of operators, heavily depend upon as close to real physics as possible \cite{7363441, 9076603}. Physical interactions can be perhaps ignored in 3D user interfaces up to a point, yet with the vast availability of mixed reality technologies on the market and increasingly more powerful hardware, the need to model physics properties has significantly increased. Object manipulation even in multi-user environments becomes more and more popular and interaction users initiate with the environment should inherently include physical properties otherwise these may be perceived as breaking immersion \cite{10.1145/1101616.1101632}.

Sadly, the application of Fitts' law towards manipulation is severely limited \cite{10.1145/97243.97278}, particularly due to being intended for pointing tasks in the first place. Yet, if we break down the phases of pointing and manipulation, we can see that these do not differ that much from another. Nieuwenhuizen \cite{5307642} studied and proposed the phases observed in 3D goal-directed movements. He proposed five phases that are generally seen when interacting with objects, latency, initiation, ballistic, correction and verification phase. During the first two phases, the velocity of hand movements is minimal, while during the ballistic phase it is at its maximum. During the correction and verification phase, velocity drops as users correct any object errors such as increasing accuracy of placements. This should not differ for either pointing or manipulation. For simplicity purposes, both pointing and manipulation can, in essence, be broken down into three at minimum parts: (a) acquisition or grasping phase, (b) transportation phase and (c) correction phase. The first phase (a) would merely differ in its name depending on either pointing or manipulation while (b) and (c) would, in essence, be similar. 

The closest work that investigated the applicability of Fitts' model for manipulation tasks is that of Yanqing and L. MacKenzie \cite{10.1145/302979.302989}. While not introducing a new model, they concluded that object size, similar to Hoffmann's model \cite{doi:10.1080/00140139508925153} shown in \autoref{eqn:hoffmann} greatly affects MT, with bigger dimensions corresponding to improved performance i.e. lesser MT and also linearly correlated with time. In \autoref{table:ModelComparisons} below, we summarize all model equations investigated thus far, including their characteristics and applicability under different spatial settings.

\begin{table*}\centering
\begin{small}
\begin{tabular}{lllccccc} \toprule
\multirow{2}{*}{\textbf{Human Performance Models}} & \multicolumn{2}{c}{Model Formulation} & \multicolumn{5}{c}{Model Characteristics} \\
      \cmidrule(lr){2-3} \cmidrule(lr){4-8}
 & MT & $ID_{t}$ / $ID_{r}$* & Based On & Space & Dir.* & Inc.** & Depth \\ \midrule
 Fitts' \cite{fitts1954information} & $MT = a + b \cdot ID$ &  $ID_{t} = log_2 \left ( \frac{2A}{W} \right )$ & N/A & 2D & No & No & No \\ \addlinespace[0.1cm]
 Shannon's \cite{10.1207/s15327051hci0701_3} & $MT = a + b \cdot ID$ &  $ID_{t} = log_2 \left ( \frac{A}{W} + 1 \right )$ & \cite{fitts1954information} & 2D & No & No & No\\ \addlinespace[0.1cm]
 Hoffmann's \cite{doi:10.1080/00140139508925153} & $MT = a + b \cdot ID$ &  $ID_{t} = log_2 \left ( \frac{2A}{W+F} \right )$ & \cite{fitts1954information} & 2D & No & No & No\\  \addlinespace[0.1cm]
 Welford's \cite{welford1968fundamentals} & $MT = a + b \cdot ID$ & $ID_{t} = log_2 \left ( \frac{A}{W} + 0.5 \right )$ & \cite{fitts1954information} & 2D & No & No & No\\ \addlinespace[0.1cm]
 Murata and Iwase's \cite{murata2001extending} & $MT = a + b \cdot ID$ &  $ID_{t} = log_2 \left ( \frac{A}{W} + 1 \right ) + c \cdot \sin{\theta}$ & \cite{10.1207/s15327051hci0701_3} & 3D & Yes & No & No\\ \addlinespace[0.1cm]
 Cha and Myung's \cite{CHA2013350} & $ MT = a + b \cdot \theta_{1} + c \cdot \sin{\theta_{2}} + d \cdot ID$ &  $ID_{t} = log_2 \left ( \frac{2A}{W+F} \right )$ &  \cite{murata2001extending, doi:10.1080/00140139508925153} & 3D & Yes & Yes & No\\ \addlinespace[0.1cm]
 Stoelen and Akin's  \cite{doi:10.1177/0018720810366560}  & $MT = a + b \cdot [ID_{t} + ID_{r}]$ &  $ID_{t/r} = log_2 \left ( \frac{A_{t} / \alpha_{r}}{W_{t} / \omega_{r}} + 1 \right )$ & \cite{10.1207/s15327051hci0701_3} & 3D & No & No & No\\ 
 \addlinespace[0.1cm]
 Machuca and Stuerzlinger's  \cite{10.1145/3290605.3300437} & $MT = a + b \cdot ID + c \cdot CTD$ &  $ID_{t} = log_2 \left ( \frac{A}{W} + 1 \right )$ & \cite{10.1207/s15327051hci0701_3} & 3D & No*** & No & Yes\\ 
\bottomrule
\end{tabular}
\end{small}
\vspace{1mm}
\caption{Summary of the most widely used 2D and 3D extensions of Fitts' law. Table illustrates the equations as defined by the respective authors. Model characteristics represent the model settings and whether these are covering important spatial characteristics to model full 3D performance. $ID_{t/r}$: Index of difficulty of translation ($t$) or rotation ($r$). Dir.*: Directions. Inc.**: Inclines. No***: Effects were investigated but no formulation or model extension was performed. [N/A]: Not applicable.}
\Description[Summary of the most widely used human performance models to date.]{This table represents the most widely used model extensions of Fitts' law, reporting the model equations and the model characteristics.}
\vspace{-2.5mm}
\label{table:ModelComparisons}
\end{table*}

\section{Outlook, Challenges and Frontiers}
To this point, we investigated and analyzed the most prominent and widely used extension of Fitts' law. However, there are still numerous challenges to address if one wants to propose a unified 3D model covering the entirety of the 3D domain. Hence in this section, we will briefly go over the challenges, frontiers and general outlook of what researchers should expect when deriving a standardized human performance metric in full 3D space. Finally, in \autoref{table:LiteratureSummary} we summarize potentially important research directions, aims and questions as the result of this review including sources and readings for other researchers to pursue.

\subsection{Influences of Depth Perception}
The main limitation of applying Fitts' model in 3D pointing, especially with the use of Mixed Reality (MR) technologies is mostly attributed to impaired depth perception \cite{doi:10.1002/jsid.303, 6798834, 8797975}. Numerous studies support that the estimation of distances for virtual targets differs to that of physical targets, in that humans appear to overestimate their ability to perceive depth in virtual environments (VEs) and by extent the target depth to reach \cite{doi:10.1177/154193129503902006, 10.1145/2543581.2543590, Witmer:1998:JPT:1246749.1246755, 7164348}. One study even estimated that this discrepancy differed with the real target to an almost 74\% of their true distance \cite{10.1145/2543581.2543590}. Other studies argued that to overcome depth limitations, one could increase the display resolution of 3D displays to provide higher representations of depth, potentially implying that this may mitigate to some extent distance overestimation \cite{8798026, kenyon2014vision}. As mentioned, Machuca and Stuerzlinger investigated and proposed a simple model accounting for depth perception in VEs. Their main finding was that movements along the depth axis i.e. away or towards the user appears to be more difficult than left to right movements, which appears to be supported by earlier work \cite{murata2001extending, CHA2013350, doi:10.1080/14640747608400584}. Including as such the effect of depth distances and adding these to a model is vital and with Machuca and Stuerzlinger's work, it does appear to have a significant effect \cite{10.1145/3290605.3300437}. Whether it shares a linear relationship as with the latter, needs more verification and testing since studies investigating the effects of depth perception on Fitts' law is still limited.

\subsection{Evaluation Approaches}
A popular approach in evaluating Fitts' model and its extensions, is to use the coefficient of determination ($R^2$) to assess the correlation between the ID and MT drawn on the x and y-axis respectively. The closer this value is to 1, the "better" the fit between these two variables and closer to 0 explaining "less" in return. Generally, an extension is deemed to be "superior" when tested against other models when there is a better correlation. However, there are numerous disadvantages to merely using the ($R^2$), which is supported by current literature \cite{10.1145/1753846.1753867, 10.1145/3173574.3173770}.

The major disadvantage of merely reporting the coefficient of determination is that it is highly dependent upon the number of data points recorded. More specifically, the more data pairs there are in the evaluation, the correlation will usually be lower. This can be exploited by having a small number of IDs and a lot of repetitions of the same tasks to achieve relatively "easily" a good resolution. 

\begin{figure*}
  \centering
  \includegraphics[width=\textwidth]{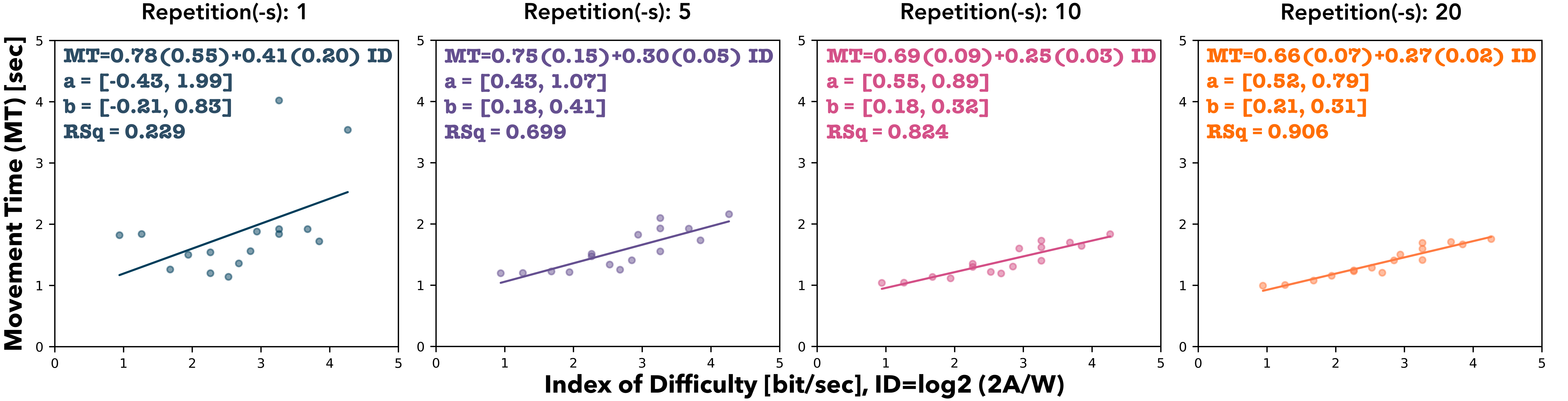}
  \vspace{-5mm}
  \caption{An example of how 1,5,10 and 20 repetitions of a simple point to the target task affect linear regression and more specifically the influence of ($R^2$). Notice the significant increase of the ($R^2$) for repetition 1, 5, 10 and 20 corresponding to $R^2=0.229$, $R^2=0.699$, $R^2=0.824$ and $R^2=0.906$ respectively. This figure serves as an example of a "more" helpful overview of the model fitting results between the ID and MT should be. More specifically we report the full line equation $MT = a (SE) + b (SE) \cdot ID$, with $(SE)$ representing the standard error and reporting the lower and upper bound confidence intervals of the constants (CI=95\%). We advise researchers to follow a similar approach instead of merely focusing on and reporting the $R^2$ values.}
  \Description[Effects of task repetitions on regression.]{The four images from left to right illustrate the influence of task repetitions on the coefficient of determination. The more data points we collect, the higher the coefficient of determination becomes. It would thus be more useful to not rely solely on the ($R^2$), but instead also provide the full line equation including standard error and standard deviation at a confidence interval ideally at 95\%.}
  \vspace{-5mm}
    ~\label{figure:regression_comparison}
\end{figure*}

Heiko Drewes \cite{10.1145/1753846.1753867} illustrated this limitation by presenting the difference of the ($R^2$) results in a single click-the-target experiment versus the same experiment but repeated multiple times and averaged over. Motivated by his observation, we took the same notion and applied it in our case as well with more levels of repetitions. We conducted a simple pointing task with four target sizes (W = 5, 7.5, 10, 12.5 [cm]) and four target separations (A = 12, 24, 36, 48 [cm]). Amounting to 16 distinctive tasks. By applying Fitts' law shown in \autoref{eqn:fitts_law_original}, a total of 16 IDs were calculated, ranging from 0.941 to 4.26 bits. A total of 20 repetitions were made by a single participant. \autoref{figure:regression_comparison} visually illustrates the reported $R^2$ values in the 3D pointing task of a single repetition versus 5, 10 and 20 repetitions. Notice that the more repetitions we have, the higher the $R^2$ value becomes. 

\begin{table*}\centering
\begin{small}
\begin{tabular}{lL{13.35cm}|rR{2.5cm}} \toprule
 \multicolumn{2}{l}{\textbf{Research Considerations and Questions to Explore}} & & \textbf{Sources \& Readings} \\ \midrule
\rowcolor{Gray} \multirow{1}{*}{\textbf{Q1}} & Explore the effect of object size as suggested by Hoffmann in the context of both pointing and manipulation with more levels and a more exhaustive setting. & &
\cite{doi:10.1080/00140139.2011.614356, 10.1145/302979.302989}  \\ 

\multirow{1}{*}{\textbf{Q2}} & In Hoffmann's formulation, can we assume that the object size (F) can represent the dimensions of a given 3D object as it does in representing the pad size index finger? Are we likely in need of different definitions? & &
\cite{doi:10.1080/00140139.2011.614356, 10.1145/302979.302989} \\ 

\rowcolor{Gray} \multirow{1}{*}{\textbf{Q3}} & How do different input technologies affect a given model? Can it be accounted for by implementing an additional term in the formulation or controlling existing terms with additional constants? & &
\cite{BURNO20154342, doi:10.1177/154193120905301216, doi:10.1177/1071181397041002119} \\ 

\multirow{1}{*}{\textbf{Q4}} & Spatial arrangements, directions \& inclinations, appear to matter according to Murata \& Iwase, Cha \& Myung and Machuca and Stuerzlinger, but can their findings be confirmed and furthermore extended in combined translational and rotational movements? Potentially of significant importance due to limited focus. & &
\cite{murata2001extending, CHA2013350, 10.1145/3290605.3300437, vaughan2010extending} \\ 

\rowcolor{Gray} \multirow{1}{*}{\textbf{Q5}} & Is the simplicity of adding two indices of difficulty of translation and rotation feasible? Can rotation indeed be modelled by merely replacing the target distance (A) with the rotational offset ($\alpha$) and the target width (W) with the rotational tolerance ($\omega$)? For translational tasks, A and W have been observed by numerous studies to share a linear relationship with MT, is that the case for the less explored rotational counterpart? Does rotation perhaps share a polynomial or even exponential relationship with MT? The aforementioned should be answered and part of these are only explored and limited to two studies by Stoelen \& Akin as well as Kulik et al., limiting their observations in movements across one line only with no directions or inclinations. & &
\cite{doi:10.1177/0018720810366560, 10.5555/2386042.2386050, 10.1145/274644.274688} \\ 

\multirow{1}{*}{\textbf{Q6}} & The above point in mind, researchers are advised to explore other extensions as well in combined movements. Both Stoelen \& Akin's, as well as Kulik's et al. work, extended Shannon's formulation exclusively. Perhaps Hoffmann's or Welford's could be explored/extended as well and their differences reported. & &
\cite{doi:10.1177/0018720810366560, 10.5555/2386042.2386050, 10.1145/274644.274688, 10.1207/s15327051hci0701_3, 10.1145/1357054.1357308, doi:10.1080/00140139.2011.614356, welford1968fundamentals} \\ 

\rowcolor{Gray} \multirow{1}{*}{\textbf{Q7}} & Depth perception in virtual environments differs from that of the real world, with users overestimating the distances. Can the work of Machuca \& Stuerzlinger, which is to the best of our knowledge the only work that accounted for the depth variable and added to Fitts' (Shannon's) extensions, suffice? With the spike in VR/AR technologies, verification of the above appears to be crucial. & &
\cite{10.1145/3290605.3300437, lin2015interaction, durgin1995comparing, Witmer:1998:JPT:1246749.1246755, doi:10.1177/154193129503902006, 10.1145/2543581.2543590, 7164348}\\ 

\multirow{1}{*}{\textbf{Q8}} & Do not evaluate models solely based on the coefficient of determination ($R^2$). In this work, we explored the limitations of purely relying on the $R^2$. Instead, aim to include the full line equation (MT) with the reported constants, including standard error and lower as well as upper bound confidence intervals (ideally at CI=95\%). & &
\cite{10.1145/1753846.1753867, 10.1145/3173574.3173770} \\ 

\rowcolor{Gray} \multirow{1}{*}{\textbf{Q9}} & Can human factors be accounted for? Tiredness and concentration for example are key elements affecting significantly human performance, yet modelling these factors is challenging and remains a gap in the literature. If this is infeasible, researchers should account for these factors during participant recruitment. & &
\cite{10.3389/fpsyg.2015.00800, 10.1145/1753846.1753867, ROZAND2015219, doi:10.1177/1071181312561207, poletti2017strategic, 10.3389/fpsyg.2018.00560} \\ 

\multirow{1}{*}{\textbf{Q10}} & Can we reduce the complexities that arise in full 3D space with the use of different sensory modalities? How do other modalities affect object interaction, in particular haptic or auditory feedback? Can we reduce potentially impaired amounts of depth perception by introducing haptic feedback in the sensory interface? & &
\cite{10.5555/2013881.2014216, 4811026, 962082, 8446227, 9076603, bunt1998multimodal, Amedi2001VisuohapticOA, TURK2014189, 10.1145/568513.568514, 10.1145/1452392.1452443, 10.3389/fpsyg.2018.00560} \\ 

\rowcolor{Gray} \multirow{1}{*}{\textbf{Q11}} & Manipulation entails different physical influences such as gravity, contact points and the mass of the object. These should prove invaluable, especially in realistic simulations and training scenarios attempting to model performance under realistic settings. Can these properties be modelled? Furthermore, how do different hand types with different contact points and degrees of freedom (e.g. grippers) and their size affect the model?  & &
\cite{9076603, 6335488, 10.1145/2858036.2858383, 10.1145/238386.238534, 10.1145/274644.274688, 10.5555/2386042.2386050, 10.3389/fpsyg.2018.00560} \\ 

\multirow{1}{*}{\textbf{Q12}} & Last but not least, what are the differences in pointing and manipulation, can these two different types of interactions be modelled under one unified formula? More importantly, can all of the aforementioned points be considered and accounted for or should we expect different formulations based on these factors alone? & & All of the above including the contents of this work. \\
\bottomrule
\end{tabular}
\end{small}
\vspace{1mm}
  \caption{The table represents potentially important directions and research paths we recommend researchers to pursue for the derivation of a robust performance metric in 3D space. The aforementioned points may prove to be crucial in solidifying such a metric and increase inter-study comparability with the hopes of generalizing it in a multitude of different settings.}
  \label{table:LiteratureSummary}
  \Description[Summary of literature review.]{In this table we summarise all the investigated literature to date attempting to extend Fitts' law in higher dimensions in addition to adding important research considerations for researchers as well as the respective sources and readings.}
\end{table*}

Unfortunately, most literature to date use this evaluation approach exclusively \cite{8797975, murata2001extending, CHA2013350, doi:10.1177/0018720810366560, 8998368}. While it does indeed provide an overview of how well the model does, researchers would also be advised to report the values of constants and their respective standard error as shown in \autoref{figure:regression_comparison}. For example, merely stating that a model has a correlation of $R^2=0.906$ is not very helpful. Instead, it would be much more useful to not only provide the latter, but also include a statement such as "Fitts' original formulation showed a correlation of $R^2=0.906$ with MT = 0.66 (0.07) [0.52, 0.79] + 0.27 (0.02) [0.21, 0.31] $\cdot$ ID". The terms represent the a and b constants respectively, reporting the standard error in addition to the lower and upper bound confidence intervals (ideally CI=95\%) that can be reported through regression analysis. 

Other methods that are also used to predict the differences between the MT and ID for a particular model and its fitting, is the Root-Mean-Square Deviation (RMSD) as well as the Mean Absolute Deviation (MAD) \cite{jastrzembski2007model}. Solely relying on the $R^2$, or for any statistical model in that case, as we saw has numerous limitations, particularly limiting inter-study comparability. Perhaps a better way for authors to support their argumentation is a mixture of multiple statistical models to show their model fitting. 

\subsection{Human Factors and Experimental Design}
By definition, Fitts' prediction model, models the performance of humans. For example, human performance is dependent upon ones own personality-related factors including but not limited to age, visual health, previous exposure to certain technologies, absorption as well as cognitive ability \cite{doi:10.1177/2041669515593028, 10.1167/iovs.02-0361, sacau_laarni_hartmann_2008, 10.1145/1228716.1228753, 10.1145/3029798.3038369}. As such, factors such as tiredness, cognitive ability, concentration, being under the influence of stimulants etc. may have a determining effect on performance. While perhaps modelling and accounting for these factors in a formulation may be significantly challenging, studies should limit or retain consistency when recruiting participants. For example, all participants in a study should ideally be evaluated if they are tired or are under the influence of certain stimulants e.g. coffee, and those who do not meet the criteria should be excluded from the analysis or initial recruitment. Sadly, most studies so far miss a clear definition of the state of the participants \cite{murata2001extending, CHA2013350, 8998368} as also suggested by Heiko Drewes \cite{10.1145/1753846.1753867}. Consequently, more focus should be given when reporting participant recruitment.

Another factor to take into account is the specific experimental design that each author considered to this date when deriving a model extension. Merely deriving a human performance model under one experimental setting is insufficient as the authors are likely in the risk of observing and tuning a model only to fit that particular study manifestation. Ultimately, if a model is only applicable for very specific application settings, its relevance can become questionable and likely at risk of losing overall generalizability. These limitations can be mitigated to some extent if one would collectively include all models as seen in \autoref{table:ModelComparisons}, in addition to any subsequent ones, and compare them under one particular 3-D experimental design setting to support their argumentation. 

\subsection{Potential Benefits of Multimodality}
Multimodal interfaces can mitigate the high complexities that inherently surround interaction in VEs. It is generally argued that human interaction with the surrounding environment is inherently multi-modal \cite{bunt1998multimodal, TURK2014189, 10.1145/568513.568514}. More specifically, manipulation by itself is multi-modal, requiring more than one modality to be active to effectively control, approach and grasp an object \cite{Billardeaat8414}.

There is strong evidence that assessing performance via Fitts' law for 3D pointing tasks appears to suffer primarily due to latency \cite{10.1145/198425.198426}, impaired depth perception \cite{6798834, 8797975}, as well as the lack of haptic (tactile) feedback \cite{10.5555/2013881.2014216, 4811026, 962082, 8446227}. The latter two appear to have a direct correlation to what seem to be the benefits of multi-modal interfaces in increasing overall perception \cite{9076603, gupta1997prototyping, 8446227, Burke:2006:CEV:1180995.1181017, Billardeaat8414, triantafyllidisrobot}. 

One study investigated 3D virtual hand pointing with the index finger in addition to incorporating vibration feedback using a 3D screen display \cite{7131735}. While disregarding depth and spatial arrangements, contrary to \cite{10.1145/3290605.3300437, murata2001extending}, they found that vibration feedback provides a reasonable addition to visual feedback, which appears to be in line with existing literature on multi-modal interfaces \cite{9076603, vibration_proximity, Murray:2003:PCT:782655.782658}. This is furthermore confirmed by numerous neuroscience studies indeed confirming that the simultaneous presence of both visual and somatosensory sensory cues is beneficial, particularly due to both modalities overlapping in the same brain region\cite{IsNeoCortexEssentiallyMultisensory, Amedi2001VisuohapticOA, JamesThomasKeithHapticActivatesVisualAreas, SathianZangaldzeVisualCortexToTactilePerception}.

However, in all cases, sensory conflict can arise when a multi-modal pipeline is unable to stimulate the senses in a synchronised way, which can be counterproductive, resulting in decreased spatial and temporal immersion, effectively nullifying the benefits \cite{popescu2002multimodal, richard1994comparison}. While conclusions are indeed difficult to draw as to which modalities directly offer higher quantifiable amounts of perception, we can indirectly measure the human performance and by proposing a standardized metric, inter-study comparisons within the domain of multimodal human-computer interaction may be feasible and increase our understanding.

\subsection{Physical Influences}
As we mentioned, physical interactions are an integral part of the manipulation process. Firstly, the effects of gravity should be evaluated. To the best of our knowledge, an experimental setting with different levels of gravity has still not been used to assess how it may affect MT and the potential influence it may pose to a model. It still remains a gap to this day \cite{10.1145/1328202.1328214}. Contact points are another aspect to consider. While there are specific methods and metrics that assess the quality of a given grasp, such as the Largest-minimum resisted wrench \cite{6335488}, it may prove to be beneficial if that can also be accounted for by Fitts' law. 

Furthermore, the grasping type users initiate with an object may prove to be a determining factor towards MT and by extent a 3D performance metric covering the manipulation domain. For example, the Southampton Hand Assessment Procedure (SHAP) is a clinically validated hand function test \cite{light_chappell_kyberd_2002} reporting six distinctive grip classifications: precision-tip, lateral, tripod, spherical, power and extension grasping. Vikash Kumar and Emanuel Todorov developed a virtual reality system for hand manipulation based upon a subset of tasks from the SHAP for evaluation purposes \cite{7363441}. However, the latter did not investigate or include Fitts' model in their evaluation. To which extent these physical properties really affect a model is hard to quantify, yet it should be pursued by researchers that aim to extend a robust performance method towards manipulation.

\subsection{Which Input Technology?}
Lastly, aggravating a unified model is the multitude of different input devices one can use. As shown by previous work, Fitts' model is heavily influenced by the input technology used, whether that includes tracked wands, a regular mouse, stylus or optical hand tracking \cite{10.1145/108844.108868, 10.1145/1357054.1357306, 10.1145/1357054.1357299, 10.1145/1978942.1979183}. Perhaps proposing a 3D performance model that would aid the community and clear the confusion as to which model is appropriate, may furthermore be aggravated by this factor alone. It should come to no surprise that numerous models should be considered as these may be specific to the input device or accounting for this limitation by adding additional constants which can, however, become problematic \cite{10.1145/3290605.3300437}. 

\section{Conclusion}
\balance
With the multitude of different performance models published in the HCI community in the last years, confusion still exists as to which formula should be used and rightly so. Yet the collective effort towards a standardized human performance metric in true full 3D space is still missing and work remains largely scattered.

In this review, we analyzed the most prominent and widely used extensions of Fitts' law and their respective contributions and limitations towards the endeavour of deriving a full 3D model. It is not as straightforward as one would think. We observed that not only including all possible spatial arrangements one would expect in full 3D space is challenging to model under one formulation, but also combining translational as well as rotational requirements in tasks is by itself not an easy approach. Furthermore, closing the gap between the discrepancies of pointing and manipulation under potentially one formulation is another very important factor.

We also went beyond the current challenges and also provided a brief outlook of future frontiers that may lay ahead when deriving a true full 3D human performance model. Factors ranging from how researchers should evaluate their work to the overlooked but yet important human factors may prove to be determining aspects for a "true" 3D performance model.

\section{Acknowledgments}
This research is supported by the EPSRC CDT in Robotics and Autonomous Systems (EP/L016834/1). We would like to thank Valentina Andries for proofreading the contents of this work.

\bibliographystyle{ACM-Reference-Format}
\bibliography{references}

%%% -*-BibTeX-*-
%%% Do NOT edit. File created by BibTeX with style
%%% ACM-Reference-Format-Journals [18-Jan-2012].

\begin{thebibliography}{97}

%%% ====================================================================
%%% NOTE TO THE USER: you can override these defaults by providing
%%% customized versions of any of these macros before the \bibliography
%%% command.  Each of them MUST provide its own final punctuation,
%%% except for \shownote{}, \showDOI{}, and \showURL{}.  The latter two
%%% do not use final punctuation, in order to avoid confusing it with
%%% the Web address.
%%%
%%% To suppress output of a particular field, define its macro to expand
%%% to an empty string, or better, \unskip, like this:
%%%
%%% \newcommand{\showDOI}[1]{\unskip}   % LaTeX syntax
%%%
%%% \def \showDOI #1{\unskip}           % plain TeX syntax
%%%
%%% ====================================================================

\ifx \showCODEN    \undefined \def \showCODEN     #1{\unskip}     \fi
\ifx \showDOI      \undefined \def \showDOI       #1{#1}\fi
\ifx \showISBNx    \undefined \def \showISBNx     #1{\unskip}     \fi
\ifx \showISBNxiii \undefined \def \showISBNxiii  #1{\unskip}     \fi
\ifx \showISSN     \undefined \def \showISSN      #1{\unskip}     \fi
\ifx \showLCCN     \undefined \def \showLCCN      #1{\unskip}     \fi
\ifx \shownote     \undefined \def \shownote      #1{#1}          \fi
\ifx \showarticletitle \undefined \def \showarticletitle #1{#1}   \fi
\ifx \showURL      \undefined \def \showURL       {\relax}        \fi
% The following commands are used for tagged output and should be
% invisible to TeX
\providecommand\bibfield[2]{#2}
\providecommand\bibinfo[2]{#2}
\providecommand\natexlab[1]{#1}
\providecommand\showeprint[2][]{arXiv:#2}

\bibitem[\protect\citeauthoryear{A.~Ghazanfar and E.~Schroeder}{A.~Ghazanfar
  and E.~Schroeder}{2006}]%
        {IsNeoCortexEssentiallyMultisensory}
\bibfield{author}{\bibinfo{person}{Asif A.~Ghazanfar} {and}
  \bibinfo{person}{Charles E.~Schroeder}.} \bibinfo{year}{2006}\natexlab{}.
\newblock \showarticletitle{Is neocortex essentially multisensory?}
\newblock \bibinfo{journal}{\emph{Trends in cognitive sciences}}
  \bibinfo{volume}{10} (\bibinfo{date}{07} \bibinfo{year}{2006}),
  \bibinfo{pages}{278--85}.
\newblock
\urldef\tempurl%
\url{https://doi.org/10.1016/j.tics.2006.04.008}
\showDOI{\tempurl}


\bibitem[\protect\citeauthoryear{Aleotti, Bottazzi, and Reggiani}{Aleotti
  et~al\mbox{.}}{2002}]%
        {vibration_proximity}
\bibfield{author}{\bibinfo{person}{Jacopo Aleotti}, \bibinfo{person}{Stefano
  Bottazzi}, {and} \bibinfo{person}{Monica Reggiani}.}
  \bibinfo{year}{2002}\natexlab{}.
\newblock \bibinfo{title}{A Multimodal User Interface for Remote Object
  Exploration in Teleoperation Systems}.
\newblock
\newblock
\urldef\tempurl%
\url{https://pdfs.semanticscholar.org/98bd/ce82196ecb7e2f8f0f1a6a480f0a2a00e80d.pdf}
\showURL{%
\tempurl}


\bibitem[\protect\citeauthoryear{Amedi, Malach, Hendler, Peled, and
  Zohary}{Amedi et~al\mbox{.}}{2001}]%
        {Amedi2001VisuohapticOA}
\bibfield{author}{\bibinfo{person}{Amir Amedi}, \bibinfo{person}{Rafael
  Malach}, \bibinfo{person}{Talma Hendler}, \bibinfo{person}{Sharon Peled},
  {and} \bibinfo{person}{Ehud Zohary}.} \bibinfo{year}{2001}\natexlab{}.
\newblock \showarticletitle{Visuo-haptic object-related activation in the
  ventral visual pathway}.
\newblock \bibinfo{journal}{\emph{Nature Neuroscience}}  \bibinfo{volume}{4}
  (\bibinfo{year}{2001}), \bibinfo{pages}{324--330}.
\newblock


\bibitem[\protect\citeauthoryear{Argelaguet and Andujar}{Argelaguet and
  Andujar}{2013}]%
        {argelaguet2013survey}
\bibfield{author}{\bibinfo{person}{Ferran Argelaguet} {and}
  \bibinfo{person}{Carlos Andujar}.} \bibinfo{year}{2013}\natexlab{}.
\newblock \showarticletitle{A survey of 3D object selection techniques for
  virtual environments}.
\newblock \bibinfo{journal}{\emph{Computers \& Graphics}} \bibinfo{volume}{37},
  \bibinfo{number}{3} (\bibinfo{year}{2013}), \bibinfo{pages}{121--136}.
\newblock


\bibitem[\protect\citeauthoryear{Barrera~Machuca and
  Stuerzlinger}{Barrera~Machuca and Stuerzlinger}{2019}]%
        {10.1145/3290605.3300437}
\bibfield{author}{\bibinfo{person}{Mayra~Donaji Barrera~Machuca} {and}
  \bibinfo{person}{Wolfgang Stuerzlinger}.} \bibinfo{year}{2019}\natexlab{}.
\newblock \showarticletitle{The Effect of Stereo Display Deficiencies on
  Virtual Hand Pointing}. In \bibinfo{booktitle}{\emph{Proceedings of the 2019
  CHI Conference on Human Factors in Computing Systems}} (Glasgow, Scotland Uk)
  \emph{(\bibinfo{series}{CHI ’19})}. \bibinfo{publisher}{Association for
  Computing Machinery}, \bibinfo{address}{New York, NY, USA}, Article
  \bibinfo{articleno}{207}, \bibinfo{numpages}{14}~pages.
\newblock
\showISBNx{9781450359702}
\urldef\tempurl%
\url{https://doi.org/10.1145/3290605.3300437}
\showDOI{\tempurl}


\bibitem[\protect\citeauthoryear{{Batmaz}, {Machuca}, {Pham}, and
  {Stuerzlinger}}{{Batmaz} et~al\mbox{.}}{2019}]%
        {8797975}
\bibfield{author}{\bibinfo{person}{A.~U. {Batmaz}}, \bibinfo{person}{M.~D.~B.
  {Machuca}}, \bibinfo{person}{D.~M. {Pham}}, {and} \bibinfo{person}{W.
  {Stuerzlinger}}.} \bibinfo{year}{2019}\natexlab{}.
\newblock \showarticletitle{Do Head-Mounted Display Stereo Deficiencies Affect
  3D Pointing Tasks in AR and VR?}. In \bibinfo{booktitle}{\emph{2019 IEEE
  Conference on Virtual Reality and 3D User Interfaces (VR)}}.
  \bibinfo{publisher}{IEEE}, \bibinfo{address}{Osaka, Japan},
  \bibinfo{pages}{585--592}.
\newblock
\urldef\tempurl%
\url{https://doi.org/10.1109/VR.2019.8797975}
\showDOI{\tempurl}


\bibitem[\protect\citeauthoryear{Bi, Li, and Zhai}{Bi et~al\mbox{.}}{2013}]%
        {10.1145/2470654.2466180}
\bibfield{author}{\bibinfo{person}{Xiaojun Bi}, \bibinfo{person}{Yang Li},
  {and} \bibinfo{person}{Shumin Zhai}.} \bibinfo{year}{2013}\natexlab{}.
\newblock \showarticletitle{FFitts Law: Modeling Finger Touch with Fitts' Law}.
  In \bibinfo{booktitle}{\emph{Proceedings of the SIGCHI Conference on Human
  Factors in Computing Systems}} (Paris, France) \emph{(\bibinfo{series}{CHI
  '13})}. \bibinfo{publisher}{Association for Computing Machinery},
  \bibinfo{address}{New York, NY, USA}, \bibinfo{pages}{1363–1372}.
\newblock
\showISBNx{9781450318990}
\urldef\tempurl%
\url{https://doi.org/10.1145/2470654.2466180}
\showDOI{\tempurl}


\bibitem[\protect\citeauthoryear{Billard and Kragic}{Billard and
  Kragic}{2019}]%
        {Billardeaat8414}
\bibfield{author}{\bibinfo{person}{Aude Billard} {and} \bibinfo{person}{Danica
  Kragic}.} \bibinfo{year}{2019}\natexlab{}.
\newblock \showarticletitle{Trends and challenges in robot manipulation}.
\newblock \bibinfo{journal}{\emph{Science}} \bibinfo{volume}{364},
  \bibinfo{number}{6446} (\bibinfo{year}{2019}), \bibinfo{pages}{1149}.
\newblock
\showISSN{0036-8075}
\urldef\tempurl%
\url{https://doi.org/10.1126/science.aat8414}
\showDOI{\tempurl}


\bibitem[\protect\citeauthoryear{{Brickler}, {Babu}, {Bertrand}, and
  {Bhargava}}{{Brickler} et~al\mbox{.}}{2018}]%
        {8446227}
\bibfield{author}{\bibinfo{person}{D. {Brickler}}, \bibinfo{person}{S.~V.
  {Babu}}, \bibinfo{person}{J. {Bertrand}}, {and} \bibinfo{person}{A.
  {Bhargava}}.} \bibinfo{year}{2018}\natexlab{}.
\newblock \showarticletitle{Towards Evaluating the Effects of Stereoscopic
  Viewing and Haptic Interaction on Perception-Action Coordination}. In
  \bibinfo{booktitle}{\emph{2018 IEEE Conference on Virtual Reality and 3D User
  Interfaces (VR)}}. \bibinfo{publisher}{IEEE}, \bibinfo{address}{Reutlingen,
  Germany}, \bibinfo{pages}{1--516}.
\newblock
\urldef\tempurl%
\url{https://doi.org/10.1109/VR.2018.8446227}
\showDOI{\tempurl}


\bibitem[\protect\citeauthoryear{Bunt, Beun, and Borghuis}{Bunt
  et~al\mbox{.}}{1998}]%
        {bunt1998multimodal}
\bibfield{author}{\bibinfo{person}{Harry Bunt}, \bibinfo{person}{Robbert-Jan
  Beun}, {and} \bibinfo{person}{Tijn Borghuis}.}
  \bibinfo{year}{1998}\natexlab{}.
\newblock \bibinfo{booktitle}{\emph{Multimodal human-computer communication:
  systems, techniques, and experiments}}. Vol.~\bibinfo{volume}{1374}.
\newblock \bibinfo{publisher}{Springer Science \& Business Media},
  \bibinfo{address}{Berlin-Heidelberg, Germany}.
\newblock


\bibitem[\protect\citeauthoryear{Burke, Prewett, Gray, Yang, Stilson, Coovert,
  Elliot, and Redden}{Burke et~al\mbox{.}}{2006}]%
        {Burke:2006:CEV:1180995.1181017}
\bibfield{author}{\bibinfo{person}{Jennifer~L. Burke},
  \bibinfo{person}{Matthew~S. Prewett}, \bibinfo{person}{Ashley~A. Gray},
  \bibinfo{person}{Liuquin Yang}, \bibinfo{person}{Frederick R.~B. Stilson},
  \bibinfo{person}{Michael~D. Coovert}, \bibinfo{person}{Linda~R. Elliot},
  {and} \bibinfo{person}{Elizabeth Redden}.} \bibinfo{year}{2006}\natexlab{}.
\newblock \showarticletitle{Comparing the Effects of Visual-auditory and
  Visual-tactile Feedback on User Performance: A Meta-analysis}. In
  \bibinfo{booktitle}{\emph{Proceedings of the 8th International Conference on
  Multimodal Interfaces}} (Banff, Alberta, Canada) \emph{(\bibinfo{series}{ICMI
  '06})}. \bibinfo{publisher}{ACM}, \bibinfo{address}{New York, NY, USA},
  \bibinfo{pages}{108--117}.
\newblock
\showISBNx{1-59593-541-X}
\urldef\tempurl%
\url{https://doi.org/10.1145/1180995.1181017}
\showDOI{\tempurl}


\bibitem[\protect\citeauthoryear{Burno, Wu, Doherty, Colett, and
  Elnaggar}{Burno et~al\mbox{.}}{2015}]%
        {BURNO20154342}
\bibfield{author}{\bibinfo{person}{Rachael~A. Burno}, \bibinfo{person}{Bing
  Wu}, \bibinfo{person}{Rina Doherty}, \bibinfo{person}{Hannah Colett}, {and}
  \bibinfo{person}{Rania Elnaggar}.} \bibinfo{year}{2015}\natexlab{}.
\newblock \showarticletitle{Applying Fitts’ Law to Gesture Based Computer
  Interactions}.
\newblock \bibinfo{journal}{\emph{Procedia Manufacturing}}  \bibinfo{volume}{3}
  (\bibinfo{year}{2015}), \bibinfo{pages}{4342 -- 4349}.
\newblock
\showISSN{2351-9789}
\urldef\tempurl%
\url{https://doi.org/10.1016/j.promfg.2015.07.429}
\showDOI{\tempurl}
\newblock
\shownote{6th International Conference on Applied Human Factors and Ergonomics
  (AHFE 2015) and the Affiliated Conferences, AHFE 2015.}


\bibitem[\protect\citeauthoryear{Burstyn, Carrascal, and Vertegaal}{Burstyn
  et~al\mbox{.}}{2016}]%
        {10.1145/2858036.2858383}
\bibfield{author}{\bibinfo{person}{Jesse Burstyn}, \bibinfo{person}{Juan~Pablo
  Carrascal}, {and} \bibinfo{person}{Roel Vertegaal}.}
  \bibinfo{year}{2016}\natexlab{}.
\newblock \showarticletitle{Fitts' Law and the Effects of Input Mapping and
  Stiffness on Flexible Display Interactions}. In
  \bibinfo{booktitle}{\emph{Proceedings of the 2016 CHI Conference on Human
  Factors in Computing Systems}} (San Jose, California, USA)
  \emph{(\bibinfo{series}{CHI '16})}. \bibinfo{publisher}{Association for
  Computing Machinery}, \bibinfo{address}{New York, NY, USA},
  \bibinfo{pages}{3649–3658}.
\newblock
\showISBNx{9781450333627}
\urldef\tempurl%
\url{https://doi.org/10.1145/2858036.2858383}
\showDOI{\tempurl}


\bibitem[\protect\citeauthoryear{Cha and Myung}{Cha and Myung}{2013}]%
        {CHA2013350}
\bibfield{author}{\bibinfo{person}{Yeonjoo Cha} {and} \bibinfo{person}{Rohae
  Myung}.} \bibinfo{year}{2013}\natexlab{}.
\newblock \showarticletitle{Extended Fitts' law for 3D pointing tasks using 3D
  target arrangements}.
\newblock \bibinfo{journal}{\emph{International Journal of Industrial
  Ergonomics}} \bibinfo{volume}{43}, \bibinfo{number}{4}
  (\bibinfo{year}{2013}), \bibinfo{pages}{350 -- 355}.
\newblock
\showISSN{0169-8141}
\urldef\tempurl%
\url{https://doi.org/10.1016/j.ergon.2013.05.005}
\showDOI{\tempurl}


\bibitem[\protect\citeauthoryear{Crossman and Goodeve}{Crossman and
  Goodeve}{1983}]%
        {doi:10.1080/14640748308402133}
\bibfield{author}{\bibinfo{person}{E.~R. F.~W. Crossman} {and}
  \bibinfo{person}{P.~J. Goodeve}.} \bibinfo{year}{1983}\natexlab{}.
\newblock \showarticletitle{Feedback control of hand-movement and Fitts' law}.
\newblock \bibinfo{journal}{\emph{The Quarterly Journal of Experimental
  Psychology Section A}} \bibinfo{volume}{35}, \bibinfo{number}{2}
  (\bibinfo{year}{1983}), \bibinfo{pages}{251--278}.
\newblock
\urldef\tempurl%
\url{https://doi.org/10.1080/14640748308402133}
\showDOI{\tempurl}


\bibitem[\protect\citeauthoryear{de~Grosbois, Heath, and Tremblay}{de~Grosbois
  et~al\mbox{.}}{2015}]%
        {10.3389/fpsyg.2015.00800}
\bibfield{author}{\bibinfo{person}{John de Grosbois}, \bibinfo{person}{Matthew
  Heath}, {and} \bibinfo{person}{Luc Tremblay}.}
  \bibinfo{year}{2015}\natexlab{}.
\newblock \showarticletitle{Augmented feedback influences upper limb reaching
  movement times but does not explain violations of Fitts' Law}.
\newblock \bibinfo{journal}{\emph{Frontiers in Psychology}}
  \bibinfo{volume}{6} (\bibinfo{year}{2015}), \bibinfo{pages}{800}.
\newblock
\showISSN{1664-1078}
\urldef\tempurl%
\url{https://doi.org/10.3389/fpsyg.2015.00800}
\showDOI{\tempurl}


\bibitem[\protect\citeauthoryear{Drewes}{Drewes}{2010}]%
        {10.1145/1753846.1753867}
\bibfield{author}{\bibinfo{person}{Heiko Drewes}.}
  \bibinfo{year}{2010}\natexlab{}.
\newblock \showarticletitle{Only One Fitts’ Law Formula Please!}. In
  \bibinfo{booktitle}{\emph{CHI ’10 Extended Abstracts on Human Factors in
  Computing Systems}} (Atlanta, Georgia, USA) \emph{(\bibinfo{series}{CHI EA
  ’10})}. \bibinfo{publisher}{Association for Computing Machinery},
  \bibinfo{address}{New York, NY, USA}, \bibinfo{pages}{2813–2822}.
\newblock
\showISBNx{9781605589305}
\urldef\tempurl%
\url{https://doi.org/10.1145/1753846.1753867}
\showDOI{\tempurl}


\bibitem[\protect\citeauthoryear{Durgin, Proffitt, Olson, and Reinke}{Durgin
  et~al\mbox{.}}{1995}]%
        {durgin1995comparing}
\bibfield{author}{\bibinfo{person}{Frank~H Durgin}, \bibinfo{person}{Dennis~R
  Proffitt}, \bibinfo{person}{Thomas~J Olson}, {and} \bibinfo{person}{Karen~S
  Reinke}.} \bibinfo{year}{1995}\natexlab{}.
\newblock \showarticletitle{Comparing depth from motion with depth from
  binocular disparity.}
\newblock \bibinfo{journal}{\emph{Journal of Experimental Psychology: Human
  Perception and Performance}} \bibinfo{volume}{21}, \bibinfo{number}{3}
  (\bibinfo{year}{1995}), \bibinfo{pages}{679}.
\newblock


\bibitem[\protect\citeauthoryear{Fitts}{Fitts}{1954}]%
        {fitts1954information}
\bibfield{author}{\bibinfo{person}{Paul~M Fitts}.}
  \bibinfo{year}{1954}\natexlab{}.
\newblock \showarticletitle{The information capacity of the human motor system
  in controlling the amplitude of movement.}
\newblock \bibinfo{journal}{\emph{Journal of experimental psychology}}
  \bibinfo{volume}{47}, \bibinfo{number}{6} (\bibinfo{year}{1954}),
  \bibinfo{pages}{381}.
\newblock


\bibitem[\protect\citeauthoryear{Fitts and Peterson}{Fitts and
  Peterson}{1964}]%
        {fitts1964information}
\bibfield{author}{\bibinfo{person}{Paul~M Fitts} {and} \bibinfo{person}{James~R
  Peterson}.} \bibinfo{year}{1964}\natexlab{}.
\newblock \showarticletitle{Information capacity of discrete motor responses.}
\newblock \bibinfo{journal}{\emph{Journal of experimental psychology}}
  \bibinfo{volume}{67}, \bibinfo{number}{2} (\bibinfo{year}{1964}),
  \bibinfo{pages}{103}.
\newblock


\bibitem[\protect\citeauthoryear{Forlines and Balakrishnan}{Forlines and
  Balakrishnan}{2008}]%
        {10.1145/1357054.1357299}
\bibfield{author}{\bibinfo{person}{Clifton Forlines} {and}
  \bibinfo{person}{Ravin Balakrishnan}.} \bibinfo{year}{2008}\natexlab{}.
\newblock \showarticletitle{Evaluating Tactile Feedback and Direct vs. Indirect
  Stylus Input in Pointing and Crossing Selection Tasks}. In
  \bibinfo{booktitle}{\emph{Proceedings of the SIGCHI Conference on Human
  Factors in Computing Systems}} (Florence, Italy) \emph{(\bibinfo{series}{CHI
  ’08})}. \bibinfo{publisher}{Association for Computing Machinery},
  \bibinfo{address}{New York, NY, USA}, \bibinfo{pages}{1563–1572}.
\newblock
\showISBNx{9781605580111}
\urldef\tempurl%
\url{https://doi.org/10.1145/1357054.1357299}
\showDOI{\tempurl}


\bibitem[\protect\citeauthoryear{{Gemperle}, {Ota}, and {Siewiorek}}{{Gemperle}
  et~al\mbox{.}}{2001}]%
        {962082}
\bibfield{author}{\bibinfo{person}{F. {Gemperle}}, \bibinfo{person}{N. {Ota}},
  {and} \bibinfo{person}{D. {Siewiorek}}.} \bibinfo{year}{2001}\natexlab{}.
\newblock \showarticletitle{Design of a wearable tactile display}. In
  \bibinfo{booktitle}{\emph{Proceedings Fifth International Symposium on
  Wearable Computers}}. \bibinfo{publisher}{IEEE}, \bibinfo{address}{Zurich,
  Switzerland}, \bibinfo{pages}{5--12}.
\newblock
\showISSN{1530-0811}
\urldef\tempurl%
\url{https://doi.org/10.1109/ISWC.2001.962082}
\showDOI{\tempurl}


\bibitem[\protect\citeauthoryear{Gillan, Holden, Adam, Rudisill, and
  Magee}{Gillan et~al\mbox{.}}{1990}]%
        {10.1145/97243.97278}
\bibfield{author}{\bibinfo{person}{Douglas~J. Gillan}, \bibinfo{person}{Kritina
  Holden}, \bibinfo{person}{Susan Adam}, \bibinfo{person}{Marianne Rudisill},
  {and} \bibinfo{person}{Laura Magee}.} \bibinfo{year}{1990}\natexlab{}.
\newblock \showarticletitle{How Does Fitts' Law Fit Pointing and Dragging?}. In
  \bibinfo{booktitle}{\emph{Proceedings of the SIGCHI Conference on Human
  Factors in Computing Systems}} (Seattle, Washington, USA)
  \emph{(\bibinfo{series}{CHI '90})}. \bibinfo{publisher}{Association for
  Computing Machinery}, \bibinfo{address}{New York, NY, USA},
  \bibinfo{pages}{227–234}.
\newblock
\showISBNx{0201509326}
\urldef\tempurl%
\url{https://doi.org/10.1145/97243.97278}
\showDOI{\tempurl}


\bibitem[\protect\citeauthoryear{Gillan, Holden, Adam, Rudisill, and
  Magee}{Gillan et~al\mbox{.}}{1992}]%
        {GILLAN1992291}
\bibfield{author}{\bibinfo{person}{Douglas~J Gillan}, \bibinfo{person}{Kritina
  Holden}, \bibinfo{person}{Susan Adam}, \bibinfo{person}{Marianne Rudisill},
  {and} \bibinfo{person}{Laura Magee}.} \bibinfo{year}{1992}\natexlab{}.
\newblock \showarticletitle{How should fitts' law be applied to human-computer
  interaction?}
\newblock \bibinfo{journal}{\emph{Interacting with Computers}}
  \bibinfo{volume}{4}, \bibinfo{number}{3} (\bibinfo{year}{1992}),
  \bibinfo{pages}{291 -- 313}.
\newblock
\showISSN{0953-5438}
\urldef\tempurl%
\url{https://doi.org/10.1016/0953-5438(92)90019-C}
\showDOI{\tempurl}


\bibitem[\protect\citeauthoryear{{Godse}, {Khadka}, and {Banic}}{{Godse}
  et~al\mbox{.}}{2019}]%
        {8798026}
\bibfield{author}{\bibinfo{person}{A. {Godse}}, \bibinfo{person}{R. {Khadka}},
  {and} \bibinfo{person}{A. {Banic}}.} \bibinfo{year}{2019}\natexlab{}.
\newblock \showarticletitle{Evaluation of Visual Perception Manipulation in
  Virtual Reality Training Environments to Improve Golf Performance}. In
  \bibinfo{booktitle}{\emph{2019 IEEE Conference on Virtual Reality and 3D User
  Interfaces (VR)}}. \bibinfo{publisher}{IEEE}, \bibinfo{address}{Osaka,
  Japan}, \bibinfo{pages}{1807--1812}.
\newblock


\bibitem[\protect\citeauthoryear{Gori, Rioul, Guiard, and Beaudouin-Lafon}{Gori
  et~al\mbox{.}}{2018}]%
        {10.1145/3173574.3173770}
\bibfield{author}{\bibinfo{person}{Julien Gori}, \bibinfo{person}{Olivier
  Rioul}, \bibinfo{person}{Yves Guiard}, {and} \bibinfo{person}{Michel
  Beaudouin-Lafon}.} \bibinfo{year}{2018}\natexlab{}.
\newblock \showarticletitle{The Perils of Confounding Factors: How Fitts' Law
  Experiments Can Lead to False Conclusions}. In
  \bibinfo{booktitle}{\emph{Proceedings of the 2018 CHI Conference on Human
  Factors in Computing Systems}} (Montreal QC, Canada)
  \emph{(\bibinfo{series}{CHI '18})}. \bibinfo{publisher}{Association for
  Computing Machinery}, \bibinfo{address}{New York, NY, USA},
  \bibinfo{pages}{1–10}.
\newblock
\showISBNx{9781450356206}
\urldef\tempurl%
\url{https://doi.org/10.1145/3173574.3173770}
\showDOI{\tempurl}


\bibitem[\protect\citeauthoryear{Grossman and Balakrishnan}{Grossman and
  Balakrishnan}{2004}]%
        {10.1145/985692.985749}
\bibfield{author}{\bibinfo{person}{Tovi Grossman} {and} \bibinfo{person}{Ravin
  Balakrishnan}.} \bibinfo{year}{2004}\natexlab{}.
\newblock \showarticletitle{Pointing at Trivariate Targets in 3D Environments}.
  In \bibinfo{booktitle}{\emph{Proceedings of the SIGCHI Conference on Human
  Factors in Computing Systems}} (Vienna, Austria) \emph{(\bibinfo{series}{CHI
  ’04})}. \bibinfo{publisher}{Association for Computing Machinery},
  \bibinfo{address}{New York, NY, USA}, \bibinfo{pages}{447–454}.
\newblock
\showISBNx{1581137028}
\urldef\tempurl%
\url{https://doi.org/10.1145/985692.985749}
\showDOI{\tempurl}


\bibitem[\protect\citeauthoryear{Gupta, Whitney, and Zeltzer}{Gupta
  et~al\mbox{.}}{1997}]%
        {gupta1997prototyping}
\bibfield{author}{\bibinfo{person}{Rakesh Gupta}, \bibinfo{person}{Daniel
  Whitney}, {and} \bibinfo{person}{David Zeltzer}.}
  \bibinfo{year}{1997}\natexlab{}.
\newblock \showarticletitle{Prototyping and design for assembly analysis using
  multimodal virtual environments}.
\newblock \bibinfo{journal}{\emph{Computer-Aided Design}} \bibinfo{volume}{29},
  \bibinfo{number}{8} (\bibinfo{year}{1997}), \bibinfo{pages}{585--597}.
\newblock


\bibitem[\protect\citeauthoryear{{Ha} and {Woo}}{{Ha} and {Woo}}{2010}]%
        {5444713}
\bibfield{author}{\bibinfo{person}{T. {Ha}} {and} \bibinfo{person}{W. {Woo}}.}
  \bibinfo{year}{2010}\natexlab{}.
\newblock \showarticletitle{An empirical evaluation of virtual hand techniques
  for 3D object manipulation in a tangible augmented reality environment}. In
  \bibinfo{booktitle}{\emph{2010 IEEE Symposium on 3D User Interfaces (3DUI)}}.
  \bibinfo{publisher}{IEEE}, \bibinfo{address}{Waltham, MA, USA},
  \bibinfo{pages}{91--98}.
\newblock
\urldef\tempurl%
\url{https://doi.org/10.1109/3DUI.2010.5444713}
\showDOI{\tempurl}


\bibitem[\protect\citeauthoryear{Hess, To, Zhou, Wang, and Cooperstock}{Hess
  et~al\mbox{.}}{2015}]%
        {doi:10.1177/2041669515593028}
\bibfield{author}{\bibinfo{person}{Robert~F. Hess}, \bibinfo{person}{Long To},
  \bibinfo{person}{Jiawei Zhou}, \bibinfo{person}{Guangyu Wang}, {and}
  \bibinfo{person}{Jeremy~R. Cooperstock}.} \bibinfo{year}{2015}\natexlab{}.
\newblock \showarticletitle{Stereo Vision: The Haves and Have-Nots}.
\newblock \bibinfo{journal}{\emph{i-Perception}} \bibinfo{volume}{6},
  \bibinfo{number}{3} (\bibinfo{year}{2015}),
  \bibinfo{pages}{2041669515593028}.
\newblock
\urldef\tempurl%
\url{https://doi.org/10.1177/2041669515593028}
\showDOI{\tempurl}


\bibitem[\protect\citeauthoryear{Hoffmann}{Hoffmann}{1991}]%
        {doi:10.1080/00140139108967324}
\bibfield{author}{\bibinfo{person}{Errol~R. Hoffmann}.}
  \bibinfo{year}{1991}\natexlab{}.
\newblock \showarticletitle{A comparison of hand and foot movement times}.
\newblock \bibinfo{journal}{\emph{Ergonomics}} \bibinfo{volume}{34},
  \bibinfo{number}{4} (\bibinfo{year}{1991}), \bibinfo{pages}{397--406}.
\newblock
\urldef\tempurl%
\url{https://doi.org/10.1080/00140139108967324}
\showDOI{\tempurl}
\newblock
\shownote{PMID: 1860460.}


\bibitem[\protect\citeauthoryear{Hoffmann}{Hoffmann}{1995}]%
        {doi:10.1080/00140139508925153}
\bibfield{author}{\bibinfo{person}{Errol~R. Hoffmann}.}
  \bibinfo{year}{1995}\natexlab{}.
\newblock \showarticletitle{Effective target tolerance in an inverted Fitts
  task}.
\newblock \bibinfo{journal}{\emph{Ergonomics}} \bibinfo{volume}{38},
  \bibinfo{number}{4} (\bibinfo{year}{1995}), \bibinfo{pages}{828--836}.
\newblock
\urldef\tempurl%
\url{https://doi.org/10.1080/00140139508925153}
\showDOI{\tempurl}


\bibitem[\protect\citeauthoryear{Hoffmann, Drury, and Romanowski}{Hoffmann
  et~al\mbox{.}}{2011}]%
        {doi:10.1080/00140139.2011.614356}
\bibfield{author}{\bibinfo{person}{Errol~R. Hoffmann},
  \bibinfo{person}{Colin~G. Drury}, {and} \bibinfo{person}{Carol~J.
  Romanowski}.} \bibinfo{year}{2011}\natexlab{}.
\newblock \showarticletitle{Performance in one-, two- and three-dimensional
  terminal aiming tasks}.
\newblock \bibinfo{journal}{\emph{Ergonomics}} \bibinfo{volume}{54},
  \bibinfo{number}{12} (\bibinfo{year}{2011}), \bibinfo{pages}{1175--1185}.
\newblock
\urldef\tempurl%
\url{https://doi.org/10.1080/00140139.2011.614356}
\showDOI{\tempurl}
\newblock
\shownote{PMID: 22103725.}


\bibitem[\protect\citeauthoryear{{Holmes}, {Charles}, {Morrow}, {McClean}, and
  {McDonough}}{{Holmes} et~al\mbox{.}}{2016}]%
        {7546014}
\bibfield{author}{\bibinfo{person}{D.~E. {Holmes}}, \bibinfo{person}{D.~K.
  {Charles}}, \bibinfo{person}{P.~J. {Morrow}}, \bibinfo{person}{S. {McClean}},
  {and} \bibinfo{person}{S.~M. {McDonough}}.} \bibinfo{year}{2016}\natexlab{}.
\newblock \showarticletitle{Using Fitt's Law to Model Arm Motion Tracked in 3D
  by a Leap Motion Controller for Virtual Reality Upper Arm Stroke
  Rehabilitation}. In \bibinfo{booktitle}{\emph{2016 IEEE 29th International
  Symposium on Computer-Based Medical Systems (CBMS)}}.
  \bibinfo{publisher}{IEEE}, \bibinfo{address}{Dublin, Ireland},
  \bibinfo{pages}{335--336}.
\newblock


\bibitem[\protect\citeauthoryear{Hong and Kang}{Hong and Kang}{2015}]%
        {doi:10.1002/jsid.303}
\bibfield{author}{\bibinfo{person}{Hyungki Hong} {and}
  \bibinfo{person}{Seok~Hyon Kang}.} \bibinfo{year}{2015}\natexlab{}.
\newblock \showarticletitle{Measurement of the lens accommodation in viewing
  stereoscopic displays}.
\newblock \bibinfo{journal}{\emph{Journal of the Society for Information
  Display}} \bibinfo{volume}{23}, \bibinfo{number}{1} (\bibinfo{year}{2015}),
  \bibinfo{pages}{19--26}.
\newblock
\urldef\tempurl%
\url{https://doi.org/10.1002/jsid.303}
\showDOI{\tempurl}


\bibitem[\protect\citeauthoryear{James, Keith~Humphrey, Gati, Servos, S~Menon,
  and Goodale}{James et~al\mbox{.}}{2002}]%
        {JamesThomasKeithHapticActivatesVisualAreas}
\bibfield{author}{\bibinfo{person}{Thomas James}, \bibinfo{person}{G
  Keith~Humphrey}, \bibinfo{person}{Sabiha Gati}, \bibinfo{person}{Philip
  Servos}, \bibinfo{person}{Ravi S~Menon}, {and} \bibinfo{person}{Melvyn
  Goodale}.} \bibinfo{year}{2002}\natexlab{}.
\newblock \showarticletitle{Haptic study of three-dimensional objects activates
  extrastriate visual areas}.
\newblock \bibinfo{journal}{\emph{Neuropsychologia}}  \bibinfo{volume}{40}
  (\bibinfo{date}{02} \bibinfo{year}{2002}), \bibinfo{pages}{1706--14}.
\newblock
\urldef\tempurl%
\url{https://doi.org/10.1016/S0028-3932(02)00017-9}
\showDOI{\tempurl}


\bibitem[\protect\citeauthoryear{Janzen, Rajendran, and Booth}{Janzen
  et~al\mbox{.}}{2016}]%
        {10.1145/2858036.2858244}
\bibfield{author}{\bibinfo{person}{Izabelle Janzen},
  \bibinfo{person}{Vasanth~K. Rajendran}, {and} \bibinfo{person}{Kellogg~S.
  Booth}.} \bibinfo{year}{2016}\natexlab{}.
\newblock \showarticletitle{Modeling the Impact of Depth on Pointing
  Performance}. In \bibinfo{booktitle}{\emph{Proceedings of the 2016 CHI
  Conference on Human Factors in Computing Systems}} (San Jose, California,
  USA) \emph{(\bibinfo{series}{CHI '16})}. \bibinfo{publisher}{Association for
  Computing Machinery}, \bibinfo{address}{New York, NY, USA},
  \bibinfo{pages}{188–199}.
\newblock
\showISBNx{9781450333627}
\urldef\tempurl%
\url{https://doi.org/10.1145/2858036.2858244}
\showDOI{\tempurl}


\bibitem[\protect\citeauthoryear{Jastrzembski and Charness}{Jastrzembski and
  Charness}{2007}]%
        {jastrzembski2007model}
\bibfield{author}{\bibinfo{person}{Tiffany~S Jastrzembski} {and}
  \bibinfo{person}{Neil Charness}.} \bibinfo{year}{2007}\natexlab{}.
\newblock \showarticletitle{The Model Human Processor and the older adult:
  Parameter estimation and validation within a mobile phone task.}
\newblock \bibinfo{journal}{\emph{Journal of experimental psychology: applied}}
  \bibinfo{volume}{13}, \bibinfo{number}{4} (\bibinfo{year}{2007}),
  \bibinfo{pages}{224}.
\newblock


\bibitem[\protect\citeauthoryear{Kenyon and Ellis}{Kenyon and Ellis}{2014}]%
        {kenyon2014vision}
\bibfield{author}{\bibinfo{person}{Robert~V Kenyon} {and}
  \bibinfo{person}{Stephen~R Ellis}.} \bibinfo{year}{2014}\natexlab{}.
\newblock \bibinfo{booktitle}{\emph{Vision, perception, and object manipulation
  in virtual environments}}.
\newblock \bibinfo{publisher}{Springer}, \bibinfo{address}{New York, USA}.
  47--70 pages.
\newblock


\bibitem[\protect\citeauthoryear{Kerr}{Kerr}{1973}]%
        {doi:10.1080/00222895.1973.10734962}
\bibfield{author}{\bibinfo{person}{Robert Kerr}.}
  \bibinfo{year}{1973}\natexlab{}.
\newblock \showarticletitle{Movement Time in an Underwater Environment}.
\newblock \bibinfo{journal}{\emph{Journal of Motor Behavior}}
  \bibinfo{volume}{5}, \bibinfo{number}{3} (\bibinfo{year}{1973}),
  \bibinfo{pages}{175--178}.
\newblock
\urldef\tempurl%
\url{https://doi.org/10.1080/00222895.1973.10734962}
\showDOI{\tempurl}
\newblock
\shownote{PMID: 23961747.}


\bibitem[\protect\citeauthoryear{KNIGHT and DAGNALL}{KNIGHT and
  DAGNALL}{1967}]%
        {doi:10.1080/00140136708930874}
\bibfield{author}{\bibinfo{person}{A.~A. KNIGHT} {and} \bibinfo{person}{P.~R.
  DAGNALL}.} \bibinfo{year}{1967}\natexlab{}.
\newblock \showarticletitle{Precision in Movements}.
\newblock \bibinfo{journal}{\emph{Ergonomics}} \bibinfo{volume}{10},
  \bibinfo{number}{3} (\bibinfo{year}{1967}), \bibinfo{pages}{321--330}.
\newblock
\urldef\tempurl%
\url{https://doi.org/10.1080/00140136708930874}
\showDOI{\tempurl}
\newblock
\shownote{PMID: 6077520.}


\bibitem[\protect\citeauthoryear{{Kondraske}}{{Kondraske}}{1994}]%
        {412031}
\bibfield{author}{\bibinfo{person}{G.~V. {Kondraske}}.}
  \bibinfo{year}{1994}\natexlab{}.
\newblock \showarticletitle{An angular motion Fitt's Law for human performance
  modeling and prediction}. In \bibinfo{booktitle}{\emph{Proceedings of 16th
  Annual International Conference of the IEEE Engineering in Medicine and
  Biology Society}}, Vol.~\bibinfo{volume}{1}. \bibinfo{publisher}{IEEE},
  \bibinfo{address}{Baltimore, MD, USA}, \bibinfo{pages}{307--308 vol.1}.
\newblock


\bibitem[\protect\citeauthoryear{Kouroupetroglou, Pino, Balmpakakis,
  Chalastanis, Golematis, Ioannou, and Koutsoumpas}{Kouroupetroglou
  et~al\mbox{.}}{2012}]%
        {10.1007/978-3-642-34182-3_2}
\bibfield{author}{\bibinfo{person}{Georgios Kouroupetroglou},
  \bibinfo{person}{Alexandros Pino}, \bibinfo{person}{Athanasios Balmpakakis},
  \bibinfo{person}{Dimitrios Chalastanis}, \bibinfo{person}{Vasileios
  Golematis}, \bibinfo{person}{Nikolaos Ioannou}, {and}
  \bibinfo{person}{Ioannis Koutsoumpas}.} \bibinfo{year}{2012}\natexlab{}.
\newblock \showarticletitle{Using Wiimote for 2D and 3D Pointing Tasks: Gesture
  Performance Evaluation}. In \bibinfo{booktitle}{\emph{Gesture and Sign
  Language in Human-Computer Interaction and Embodied Communication}}.
  \bibinfo{publisher}{Springer Berlin Heidelberg}, \bibinfo{address}{Berlin,
  Heidelberg}, \bibinfo{pages}{13--23}.
\newblock
\showISBNx{978-3-642-34182-3}


\bibitem[\protect\citeauthoryear{{Kulik}, {Kunert}, and {Froehlich}}{{Kulik}
  et~al\mbox{.}}{2020}]%
        {8998368}
\bibfield{author}{\bibinfo{person}{A. {Kulik}}, \bibinfo{person}{A. {Kunert}},
  {and} \bibinfo{person}{B. {Froehlich}}.} \bibinfo{year}{2020}\natexlab{}.
\newblock \showarticletitle{On Motor Performance in Virtual 3D Object
  Manipulation}.
\newblock \bibinfo{journal}{\emph{IEEE Transactions on Visualization and
  Computer Graphics}} \bibinfo{volume}{26}, \bibinfo{number}{5}
  (\bibinfo{year}{2020}), \bibinfo{pages}{2041--2050}.
\newblock


\bibitem[\protect\citeauthoryear{{Kumar} and {Todorov}}{{Kumar} and
  {Todorov}}{2015}]%
        {7363441}
\bibfield{author}{\bibinfo{person}{V. {Kumar}} {and} \bibinfo{person}{E.
  {Todorov}}.} \bibinfo{year}{2015}\natexlab{}.
\newblock \showarticletitle{MuJoCo HAPTIX: A virtual reality system for hand
  manipulation}. In \bibinfo{booktitle}{\emph{2015 IEEE-RAS 15th International
  Conference on Humanoid Robots (Humanoids)}}. \bibinfo{publisher}{IEEE},
  \bibinfo{address}{Seoul, South Korea}, \bibinfo{pages}{657--663}.
\newblock
\urldef\tempurl%
\url{https://doi.org/10.1109/HUMANOIDS.2015.7363441}
\showDOI{\tempurl}


\bibitem[\protect\citeauthoryear{Kunert, Kulik, Huckauf, and
  Fr\"{o}hlich}{Kunert et~al\mbox{.}}{2007}]%
        {10.5555/2386042.2386050}
\bibfield{author}{\bibinfo{person}{Andr\'{e} Kunert},
  \bibinfo{person}{Alexander Kulik}, \bibinfo{person}{Anke Huckauf}, {and}
  \bibinfo{person}{Bernd Fr\"{o}hlich}.} \bibinfo{year}{2007}\natexlab{}.
\newblock \showarticletitle{A Comparison of Tracking- and Controller-Based
  Input for Complex Bimanual Interaction in Virtual Environments}. In
  \bibinfo{booktitle}{\emph{Proceedings of the 13th Eurographics Conference on
  Virtual Environments}} (Weimar, Germany)
  \emph{(\bibinfo{series}{EGVE’07})}. \bibinfo{publisher}{Eurographics
  Association}, \bibinfo{address}{Goslar, DEU}, \bibinfo{pages}{43–52}.
\newblock
\showISBNx{9783905674026}


\bibitem[\protect\citeauthoryear{Lampton, McDonald, Singer, and Bliss}{Lampton
  et~al\mbox{.}}{1995}]%
        {doi:10.1177/154193129503902006}
\bibfield{author}{\bibinfo{person}{Donald~R. Lampton},
  \bibinfo{person}{Daniel~P. McDonald}, \bibinfo{person}{Michael Singer}, {and}
  \bibinfo{person}{James~P. Bliss}.} \bibinfo{year}{1995}\natexlab{}.
\newblock \showarticletitle{Distance Estimation in Virtual Environments}.
\newblock \bibinfo{journal}{\emph{Proceedings of the Human Factors and
  Ergonomics Society Annual Meeting}} \bibinfo{volume}{39},
  \bibinfo{number}{20} (\bibinfo{year}{1995}), \bibinfo{pages}{1268--1272}.
\newblock
\urldef\tempurl%
\url{https://doi.org/10.1177/154193129503902006}
\showDOI{\tempurl}


\bibitem[\protect\citeauthoryear{Light, Chappell, and Kyberd}{Light
  et~al\mbox{.}}{2002}]%
        {light_chappell_kyberd_2002}
\bibfield{author}{\bibinfo{person}{Colin~M. Light}, \bibinfo{person}{Paul~H.
  Chappell}, {and} \bibinfo{person}{Peter~J. Kyberd}.}
  \bibinfo{year}{2002}\natexlab{}.
\newblock \showarticletitle{Establishing a standardized clinical assessment
  tool of pathologic and prosthetic hand function: Normative data, reliability,
  and validity}.
\newblock \bibinfo{journal}{\emph{Archives of Physical Medicine and
  Rehabilitation}} \bibinfo{volume}{83}, \bibinfo{number}{6}
  (\bibinfo{year}{2002}), \bibinfo{pages}{776–783}.
\newblock
\urldef\tempurl%
\url{https://doi.org/10.1053/apmr.2002.32737}
\showDOI{\tempurl}


\bibitem[\protect\citeauthoryear{Lin and Woldegiorgis}{Lin and
  Woldegiorgis}{2015}]%
        {lin2015interaction}
\bibfield{author}{\bibinfo{person}{Chiuhsiang~Joe Lin} {and}
  \bibinfo{person}{Bereket~Haile Woldegiorgis}.}
  \bibinfo{year}{2015}\natexlab{}.
\newblock \showarticletitle{Interaction and visual performance in stereoscopic
  displays: A review}.
\newblock \bibinfo{journal}{\emph{Journal of the Society for Information
  Display}} \bibinfo{volume}{23}, \bibinfo{number}{7} (\bibinfo{year}{2015}),
  \bibinfo{pages}{319--332}.
\newblock


\bibitem[\protect\citeauthoryear{{Liu}, {van Liere}, {Nieuwenhuizen}, and
  {Martens}}{{Liu} et~al\mbox{.}}{2009}]%
        {4811026}
\bibfield{author}{\bibinfo{person}{L. {Liu}}, \bibinfo{person}{R. {van Liere}},
  \bibinfo{person}{C. {Nieuwenhuizen}}, {and} \bibinfo{person}{J. {Martens}}.}
  \bibinfo{year}{2009}\natexlab{}.
\newblock \showarticletitle{Comparing Aimed Movements in the Real World and in
  Virtual Reality}. In \bibinfo{booktitle}{\emph{2009 IEEE Virtual Reality
  Conference}}. \bibinfo{publisher}{IEEE}, \bibinfo{address}{Lafayette, LA,
  USA}, \bibinfo{pages}{219--222}.
\newblock


\bibitem[\protect\citeauthoryear{{Lubos}, {Bruder}, and {Steinicke}}{{Lubos}
  et~al\mbox{.}}{2014}]%
        {6798834}
\bibfield{author}{\bibinfo{person}{P. {Lubos}}, \bibinfo{person}{G. {Bruder}},
  {and} \bibinfo{person}{F. {Steinicke}}.} \bibinfo{year}{2014}\natexlab{}.
\newblock \showarticletitle{Analysis of direct selection in head-mounted
  display environments}. In \bibinfo{booktitle}{\emph{2014 IEEE Symposium on 3D
  User Interfaces (3DUI)}}. \bibinfo{publisher}{IEEE},
  \bibinfo{address}{Minneapolis, MN, USA}, \bibinfo{pages}{11--18}.
\newblock


\bibitem[\protect\citeauthoryear{MacKenzie}{MacKenzie}{1992}]%
        {10.1207/s15327051hci0701_3}
\bibfield{author}{\bibinfo{person}{I.~Scott MacKenzie}.}
  \bibinfo{year}{1992}\natexlab{}.
\newblock \showarticletitle{Fitts’ Law as a Research and Design Tool in
  Human-Computer Interaction}.
\newblock \bibinfo{journal}{\emph{Hum.-Comput. Interact.}} \bibinfo{volume}{7},
  \bibinfo{number}{1} (\bibinfo{date}{March} \bibinfo{year}{1992}),
  \bibinfo{pages}{91–139}.
\newblock
\showISSN{0737-0024}
\urldef\tempurl%
\url{https://doi.org/10.1207/s15327051hci0701_3}
\showDOI{\tempurl}


\bibitem[\protect\citeauthoryear{MacKenzie and Isokoski}{MacKenzie and
  Isokoski}{2008}]%
        {10.1145/1357054.1357308}
\bibfield{author}{\bibinfo{person}{I.~Scott MacKenzie} {and}
  \bibinfo{person}{Poika Isokoski}.} \bibinfo{year}{2008}\natexlab{}.
\newblock \showarticletitle{Fitts’ Throughput and the Speed-Accuracy
  Tradeoff}. In \bibinfo{booktitle}{\emph{Proceedings of the SIGCHI Conference
  on Human Factors in Computing Systems}} (Florence, Italy)
  \emph{(\bibinfo{series}{CHI ’08})}. \bibinfo{publisher}{Association for
  Computing Machinery}, \bibinfo{address}{New York, NY, USA},
  \bibinfo{pages}{1633–1636}.
\newblock
\showISBNx{9781605580111}
\urldef\tempurl%
\url{https://doi.org/10.1145/1357054.1357308}
\showDOI{\tempurl}


\bibitem[\protect\citeauthoryear{MacKenzie, Sellen, and Buxton}{MacKenzie
  et~al\mbox{.}}{1991}]%
        {10.1145/108844.108868}
\bibfield{author}{\bibinfo{person}{I.~Scott MacKenzie},
  \bibinfo{person}{Abigail Sellen}, {and} \bibinfo{person}{William A.~S.
  Buxton}.} \bibinfo{year}{1991}\natexlab{}.
\newblock \showarticletitle{A Comparison of Input Devices in Element Pointing
  and Dragging Tasks}. In \bibinfo{booktitle}{\emph{Proceedings of the SIGCHI
  Conference on Human Factors in Computing Systems}} (New Orleans, Louisiana,
  USA) \emph{(\bibinfo{series}{CHI ’91})}. \bibinfo{publisher}{Association
  for Computing Machinery}, \bibinfo{address}{New York, NY, USA},
  \bibinfo{pages}{161–166}.
\newblock
\showISBNx{0897913833}
\urldef\tempurl%
\url{https://doi.org/10.1145/108844.108868}
\showDOI{\tempurl}


\bibitem[\protect\citeauthoryear{McGee, Amento, Brooks, and Harley}{McGee
  et~al\mbox{.}}{1997}]%
        {doi:10.1177/1071181397041002119}
\bibfield{author}{\bibinfo{person}{Mike McGee}, \bibinfo{person}{Brian Amento},
  \bibinfo{person}{Patrick Brooks}, {and} \bibinfo{person}{Hope Harley}.}
  \bibinfo{year}{1997}\natexlab{}.
\newblock \showarticletitle{Fitts and VR: Evaluating Display and Input Devices
  with Fitts' Law}.
\newblock \bibinfo{journal}{\emph{Proceedings of the Human Factors and
  Ergonomics Society Annual Meeting}} \bibinfo{volume}{41}, \bibinfo{number}{2}
  (\bibinfo{year}{1997}), \bibinfo{pages}{1259--1262}.
\newblock
\urldef\tempurl%
\url{https://doi.org/10.1177/1071181397041002119}
\showDOI{\tempurl}


\bibitem[\protect\citeauthoryear{McGlynn and Rogers}{McGlynn and
  Rogers}{2017}]%
        {10.1145/3029798.3038369}
\bibfield{author}{\bibinfo{person}{Sean~A. McGlynn} {and}
  \bibinfo{person}{Wendy~A. Rogers}.} \bibinfo{year}{2017}\natexlab{}.
\newblock \showarticletitle{Considerations for Presence in Teleoperation}. In
  \bibinfo{booktitle}{\emph{Proceedings of the Companion of the 2017 ACM/IEEE
  International Conference on Human-Robot Interaction}} (Vienna, Austria)
  \emph{(\bibinfo{series}{HRI '17})}. \bibinfo{publisher}{Association for
  Computing Machinery}, \bibinfo{address}{New York, NY, USA},
  \bibinfo{pages}{203–204}.
\newblock
\showISBNx{9781450348850}
\urldef\tempurl%
\url{https://doi.org/10.1145/3029798.3038369}
\showDOI{\tempurl}


\bibitem[\protect\citeauthoryear{Menchaca-Brandan, Liu, Oman, and
  Natapoff}{Menchaca-Brandan et~al\mbox{.}}{2007}]%
        {10.1145/1228716.1228753}
\bibfield{author}{\bibinfo{person}{M.~Alejandra Menchaca-Brandan},
  \bibinfo{person}{Andrew~M. Liu}, \bibinfo{person}{Charles~M. Oman}, {and}
  \bibinfo{person}{Alan Natapoff}.} \bibinfo{year}{2007}\natexlab{}.
\newblock \showarticletitle{Influence of Perspective-Taking and Mental Rotation
  Abilities in Space Teleoperation}. In \bibinfo{booktitle}{\emph{Proceedings
  of the ACM/IEEE International Conference on Human-Robot Interaction}}
  (Arlington, Virginia, USA) \emph{(\bibinfo{series}{HRI '07})}.
  \bibinfo{publisher}{Association for Computing Machinery},
  \bibinfo{address}{New York, NY, USA}, \bibinfo{pages}{271–278}.
\newblock
\showISBNx{9781595936172}
\urldef\tempurl%
\url{https://doi.org/10.1145/1228716.1228753}
\showDOI{\tempurl}


\bibitem[\protect\citeauthoryear{Meyer, Abrams, Kornblum, Wright, and
  Keith~Smith}{Meyer et~al\mbox{.}}{1988}]%
        {meyer1988optimality}
\bibfield{author}{\bibinfo{person}{David~E Meyer}, \bibinfo{person}{Richard~A
  Abrams}, \bibinfo{person}{Sylvan Kornblum}, \bibinfo{person}{Charles~E
  Wright}, {and} \bibinfo{person}{JE Keith~Smith}.}
  \bibinfo{year}{1988}\natexlab{}.
\newblock \showarticletitle{Optimality in human motor performance: ideal
  control of rapid aimed movements.}
\newblock \bibinfo{journal}{\emph{Psychological review}} \bibinfo{volume}{95},
  \bibinfo{number}{3} (\bibinfo{year}{1988}), \bibinfo{pages}{340}.
\newblock


\bibitem[\protect\citeauthoryear{Murata and Iwase}{Murata and Iwase}{2001}]%
        {murata2001extending}
\bibfield{author}{\bibinfo{person}{Atsuo Murata} {and}
  \bibinfo{person}{Hirokazu Iwase}.} \bibinfo{year}{2001}\natexlab{}.
\newblock \showarticletitle{Extending Fitts' law to a three-dimensional
  pointing task}.
\newblock \bibinfo{journal}{\emph{Human movement science}}
  \bibinfo{volume}{20}, \bibinfo{number}{6} (\bibinfo{year}{2001}),
  \bibinfo{pages}{791--805}.
\newblock


\bibitem[\protect\citeauthoryear{Murray, Klatzky, and Khosla}{Murray
  et~al\mbox{.}}{2003}]%
        {Murray:2003:PCT:782655.782658}
\bibfield{author}{\bibinfo{person}{Anne~M. Murray}, \bibinfo{person}{Roberta~L.
  Klatzky}, {and} \bibinfo{person}{Pradeep~K. Khosla}.}
  \bibinfo{year}{2003}\natexlab{}.
\newblock \showarticletitle{Psychophysical Characterization and Testbed
  Validation of a Wearable Vibrotactile Glove for Telemanipulation}.
\newblock \bibinfo{journal}{\emph{Presence: Teleoper. Virtual Environ.}}
  \bibinfo{volume}{12}, \bibinfo{number}{2} (\bibinfo{date}{April}
  \bibinfo{year}{2003}), \bibinfo{pages}{156--182}.
\newblock
\showISSN{1054-7460}
\urldef\tempurl%
\url{https://doi.org/10.1162/105474603321640923}
\showDOI{\tempurl}


\bibitem[\protect\citeauthoryear{Narayan, Waugh, Zhang, Bafna, and
  Bowman}{Narayan et~al\mbox{.}}{2005}]%
        {10.1145/1101616.1101632}
\bibfield{author}{\bibinfo{person}{Michael Narayan}, \bibinfo{person}{Leo
  Waugh}, \bibinfo{person}{Xiaoyu Zhang}, \bibinfo{person}{Pradyut Bafna},
  {and} \bibinfo{person}{Doug Bowman}.} \bibinfo{year}{2005}\natexlab{}.
\newblock \showarticletitle{Quantifying the Benefits of Immersion for
  Collaboration in Virtual Environments}. In
  \bibinfo{booktitle}{\emph{Proceedings of the ACM Symposium on Virtual Reality
  Software and Technology}} (Monterey, CA, USA) \emph{(\bibinfo{series}{VRST
  '05})}. \bibinfo{publisher}{Association for Computing Machinery},
  \bibinfo{address}{New York, NY, USA}, \bibinfo{pages}{78–81}.
\newblock
\showISBNx{1595930981}
\urldef\tempurl%
\url{https://doi.org/10.1145/1101616.1101632}
\showDOI{\tempurl}


\bibitem[\protect\citeauthoryear{{Nieuwenhuizen}, {Liu}, v.~{Liere}, and
  {Martens}}{{Nieuwenhuizen} et~al\mbox{.}}{2009}]%
        {5307642}
\bibfield{author}{\bibinfo{person}{K. {Nieuwenhuizen}}, \bibinfo{person}{L.
  {Liu}}, \bibinfo{person}{R. v. {Liere}}, {and} \bibinfo{person}{J.
  {Martens}}.} \bibinfo{year}{2009}\natexlab{}.
\newblock \showarticletitle{Insights from Dividing 3D Goal-Directed Movements
  into Meaningful Phases}.
\newblock \bibinfo{journal}{\emph{IEEE Computer Graphics and Applications}}
  \bibinfo{volume}{29}, \bibinfo{number}{6} (\bibinfo{year}{2009}),
  \bibinfo{pages}{44--53}.
\newblock


\bibitem[\protect\citeauthoryear{{Pfeiffer} and {Stuerzlinger}}{{Pfeiffer} and
  {Stuerzlinger}}{2015}]%
        {7131735}
\bibfield{author}{\bibinfo{person}{M. {Pfeiffer}} {and} \bibinfo{person}{W.
  {Stuerzlinger}}.} \bibinfo{year}{2015}\natexlab{}.
\newblock \showarticletitle{3D virtual hand pointing with EMS and vibration
  feedback}. In \bibinfo{booktitle}{\emph{2015 IEEE Symposium on 3D User
  Interfaces (3DUI)}}. \bibinfo{publisher}{IEEE}, \bibinfo{address}{Arles,
  France}, \bibinfo{pages}{117--120}.
\newblock


\bibitem[\protect\citeauthoryear{Pino, Tzemis, Ioannou, and
  Kouroupetroglou}{Pino et~al\mbox{.}}{2013}]%
        {10.1007/978-3-642-39330-3_38}
\bibfield{author}{\bibinfo{person}{Alexandros Pino}, \bibinfo{person}{Evangelos
  Tzemis}, \bibinfo{person}{Nikolaos Ioannou}, {and} \bibinfo{person}{Georgios
  Kouroupetroglou}.} \bibinfo{year}{2013}\natexlab{}.
\newblock \showarticletitle{Using Kinect for 2D and 3D Pointing Tasks:
  Performance Evaluation}. In \bibinfo{booktitle}{\emph{Proceedings of the 15th
  International Conference on Human-Computer Interaction: Interaction
  Modalities and Techniques - Volume Part IV}} (Las Vegas, NV)
  \emph{(\bibinfo{series}{HCI’13})}. \bibinfo{publisher}{Springer-Verlag},
  \bibinfo{address}{Berlin, Heidelberg}, \bibinfo{pages}{358–367}.
\newblock
\showISBNx{9783642393297}
\urldef\tempurl%
\url{https://doi.org/10.1007/978-3-642-39330-3_38}
\showDOI{\tempurl}


\bibitem[\protect\citeauthoryear{Poletti, Sleimen-Malkoun, Decker, Retornaz,
  Lemaire, and Temprado}{Poletti et~al\mbox{.}}{2017}]%
        {poletti2017strategic}
\bibfield{author}{\bibinfo{person}{C{\'e}line Poletti}, \bibinfo{person}{Rita
  Sleimen-Malkoun}, \bibinfo{person}{Leslie~Marion Decker},
  \bibinfo{person}{Fr{\'e}d{\'e}rique Retornaz}, \bibinfo{person}{Patrick
  Lemaire}, {and} \bibinfo{person}{Jean-Jacques Temprado}.}
  \bibinfo{year}{2017}\natexlab{}.
\newblock \showarticletitle{Strategic variations in Fitts’ task: comparison
  of healthy older adults and cognitively impaired patients}.
\newblock \bibinfo{journal}{\emph{Frontiers in aging neuroscience}}
  \bibinfo{volume}{8} (\bibinfo{year}{2017}), \bibinfo{pages}{334}.
\newblock


\bibitem[\protect\citeauthoryear{Popescu, Burdea, and Trefftz}{Popescu
  et~al\mbox{.}}{2002}]%
        {popescu2002multimodal}
\bibfield{author}{\bibinfo{person}{George~V. Popescu},
  \bibinfo{person}{Grigore~C. Burdea}, {and} \bibinfo{person}{Helmuth
  Trefftz}.} \bibinfo{year}{2002}\natexlab{}.
\newblock \bibinfo{booktitle}{\emph{Multimodal interaction modeling.}}
\newblock \bibinfo{publisher}{Lawrence Erlbaum Associates Publishers},
  \bibinfo{address}{Mahwah, NJ, US}, \bibinfo{pages}{435--454}.
\newblock
\showISBNx{0-8058-3270-X (Hardcover)}


\bibitem[\protect\citeauthoryear{Quek, McNeill, Bryll, Duncan, Ma, Kirbas,
  McCullough, and Ansari}{Quek et~al\mbox{.}}{2002}]%
        {10.1145/568513.568514}
\bibfield{author}{\bibinfo{person}{Francis Quek}, \bibinfo{person}{David
  McNeill}, \bibinfo{person}{Robert Bryll}, \bibinfo{person}{Susan Duncan},
  \bibinfo{person}{Xin-Feng Ma}, \bibinfo{person}{Cemil Kirbas},
  \bibinfo{person}{Karl~E. McCullough}, {and} \bibinfo{person}{Rashid Ansari}.}
  \bibinfo{year}{2002}\natexlab{}.
\newblock \showarticletitle{Multimodal Human Discourse: Gesture and Speech}.
\newblock \bibinfo{journal}{\emph{ACM Trans. Comput.-Hum. Interact.}}
  \bibinfo{volume}{9}, \bibinfo{number}{3} (\bibinfo{date}{Sept.}
  \bibinfo{year}{2002}), \bibinfo{pages}{171–193}.
\newblock
\showISSN{1073-0516}
\urldef\tempurl%
\url{https://doi.org/10.1145/568513.568514}
\showDOI{\tempurl}


\bibitem[\protect\citeauthoryear{Renner, Velichkovsky, and Helmert}{Renner
  et~al\mbox{.}}{2013}]%
        {10.1145/2543581.2543590}
\bibfield{author}{\bibinfo{person}{Rebekka~S. Renner},
  \bibinfo{person}{Boris~M. Velichkovsky}, {and} \bibinfo{person}{Jens~R.
  Helmert}.} \bibinfo{year}{2013}\natexlab{}.
\newblock \showarticletitle{The Perception of Egocentric Distances in Virtual
  Environments - A Review}.
\newblock \bibinfo{journal}{\emph{ACM Comput. Surv.}} \bibinfo{volume}{46},
  \bibinfo{number}{2}, Article \bibinfo{articleno}{23} (\bibinfo{date}{Dec.}
  \bibinfo{year}{2013}), \bibinfo{numpages}{40}~pages.
\newblock
\showISSN{0360-0300}
\urldef\tempurl%
\url{https://doi.org/10.1145/2543581.2543590}
\showDOI{\tempurl}


\bibitem[\protect\citeauthoryear{Richard, Burdea, Gomez, and Coiffet}{Richard
  et~al\mbox{.}}{1994}]%
        {richard1994comparison}
\bibfield{author}{\bibinfo{person}{Paul Richard}, \bibinfo{person}{Grigore
  Burdea}, \bibinfo{person}{Daniel Gomez}, {and} \bibinfo{person}{Philippe
  Coiffet}.} \bibinfo{year}{1994}\natexlab{}.
\newblock \bibinfo{title}{A comparison of haptic, visual and auditive force
  feedback for deformable virtual objects}.
\newblock
\newblock


\bibitem[\protect\citeauthoryear{Rozand, Lebon, Papaxanthis, and Lepers}{Rozand
  et~al\mbox{.}}{2015}]%
        {ROZAND2015219}
\bibfield{author}{\bibinfo{person}{V. Rozand}, \bibinfo{person}{F. Lebon},
  \bibinfo{person}{C. Papaxanthis}, {and} \bibinfo{person}{R. Lepers}.}
  \bibinfo{year}{2015}\natexlab{}.
\newblock \showarticletitle{Effect of mental fatigue on speed–accuracy
  trade-off}.
\newblock \bibinfo{journal}{\emph{Neuroscience}}  \bibinfo{volume}{297}
  (\bibinfo{year}{2015}), \bibinfo{pages}{219 -- 230}.
\newblock
\showISSN{0306-4522}
\urldef\tempurl%
\url{https://doi.org/10.1016/j.neuroscience.2015.03.066}
\showDOI{\tempurl}


\bibitem[\protect\citeauthoryear{Sacau, Laarni, and Hartmann}{Sacau
  et~al\mbox{.}}{2008}]%
        {sacau_laarni_hartmann_2008}
\bibfield{author}{\bibinfo{person}{Ana Sacau}, \bibinfo{person}{Jari Laarni},
  {and} \bibinfo{person}{Tilo Hartmann}.} \bibinfo{year}{2008}\natexlab{}.
\newblock \showarticletitle{Influence of individual factors on presence}.
\newblock \bibinfo{journal}{\emph{Computers in Human Behavior}}
  \bibinfo{volume}{24}, \bibinfo{number}{5} (\bibinfo{year}{2008}),
  \bibinfo{pages}{2255–2273}.
\newblock
\urldef\tempurl%
\url{https://doi.org/10.1016/j.chb.2007.11.001}
\showDOI{\tempurl}


\bibitem[\protect\citeauthoryear{Sasangohar, MacKenzie, and Scott}{Sasangohar
  et~al\mbox{.}}{2009}]%
        {doi:10.1177/154193120905301216}
\bibfield{author}{\bibinfo{person}{Farzan Sasangohar},
  \bibinfo{person}{I.~Scott MacKenzie}, {and} \bibinfo{person}{Stacey~D.
  Scott}.} \bibinfo{year}{2009}\natexlab{}.
\newblock \showarticletitle{Evaluation of Mouse and Touch Input for a Tabletop
  Display Using Fitts' Reciprocal Tapping Task}.
\newblock \bibinfo{journal}{\emph{Proceedings of the Human Factors and
  Ergonomics Society Annual Meeting}} \bibinfo{volume}{53},
  \bibinfo{number}{12} (\bibinfo{year}{2009}), \bibinfo{pages}{839--843}.
\newblock
\urldef\tempurl%
\url{https://doi.org/10.1177/154193120905301216}
\showDOI{\tempurl}


\bibitem[\protect\citeauthoryear{Sathian and Zangaladze}{Sathian and
  Zangaladze}{2002}]%
        {SathianZangaldzeVisualCortexToTactilePerception}
\bibfield{author}{\bibinfo{person}{K Sathian} {and} \bibinfo{person}{A
  Zangaladze}.} \bibinfo{year}{2002}\natexlab{}.
\newblock \showarticletitle{Feeling with the mind's eye: Contribution of visual
  cortex to tactile perception}.
\newblock \bibinfo{journal}{\emph{Behavioural brain research}}
  \bibinfo{volume}{135} (\bibinfo{date}{10} \bibinfo{year}{2002}),
  \bibinfo{pages}{127--32}.
\newblock
\urldef\tempurl%
\url{https://doi.org/10.1016/S0166-4328(02)00141-9}
\showDOI{\tempurl}


\bibitem[\protect\citeauthoryear{Schofield}{Schofield}{1976}]%
        {doi:10.1080/14640747608400584}
\bibfield{author}{\bibinfo{person}{W.~N. Schofield}.}
  \bibinfo{year}{1976}\natexlab{}.
\newblock \showarticletitle{Do Children Find Movements Which Cross the Body
  Midline Difficult?}
\newblock \bibinfo{journal}{\emph{Quarterly Journal of Experimental
  Psychology}} \bibinfo{volume}{28}, \bibinfo{number}{4}
  (\bibinfo{year}{1976}), \bibinfo{pages}{571--582}.
\newblock
\urldef\tempurl%
\url{https://doi.org/10.1080/14640747608400584}
\showDOI{\tempurl}


\bibitem[\protect\citeauthoryear{Shoemaker, Tsukitani, Kitamura, and
  Booth}{Shoemaker et~al\mbox{.}}{2012}]%
        {10.1145/2395131.2395135}
\bibfield{author}{\bibinfo{person}{Garth Shoemaker}, \bibinfo{person}{Takayuki
  Tsukitani}, \bibinfo{person}{Yoshifumi Kitamura}, {and}
  \bibinfo{person}{Kellogg~S. Booth}.} \bibinfo{year}{2012}\natexlab{}.
\newblock \showarticletitle{Two-Part Models Capture the Impact of Gain on
  Pointing Performance}.
\newblock \bibinfo{journal}{\emph{ACM Trans. Comput.-Hum. Interact.}}
  \bibinfo{volume}{19}, \bibinfo{number}{4}, Article \bibinfo{articleno}{28}
  (\bibinfo{date}{Dec.} \bibinfo{year}{2012}), \bibinfo{numpages}{34}~pages.
\newblock
\showISSN{1073-0516}
\urldef\tempurl%
\url{https://doi.org/10.1145/2395131.2395135}
\showDOI{\tempurl}


\bibitem[\protect\citeauthoryear{So and Griffin}{So and Griffin}{2000}]%
        {so2000effects}
\bibfield{author}{\bibinfo{person}{Richard~HY So} {and}
  \bibinfo{person}{Michael~J Griffin}.} \bibinfo{year}{2000}\natexlab{}.
\newblock \showarticletitle{Effects of a target movement direction cue on
  head-tracking performance}.
\newblock \bibinfo{journal}{\emph{Ergonomics}} \bibinfo{volume}{43},
  \bibinfo{number}{3} (\bibinfo{year}{2000}), \bibinfo{pages}{360--376}.
\newblock


\bibitem[\protect\citeauthoryear{Stoelen and Akin}{Stoelen and Akin}{2010}]%
        {doi:10.1177/0018720810366560}
\bibfield{author}{\bibinfo{person}{Martin~F. Stoelen} {and}
  \bibinfo{person}{David~L. Akin}.} \bibinfo{year}{2010}\natexlab{}.
\newblock \showarticletitle{Assessment of Fitts’ Law for Quantifying Combined
  Rotational and Translational Movements}.
\newblock \bibinfo{journal}{\emph{Human Factors}} \bibinfo{volume}{52},
  \bibinfo{number}{1} (\bibinfo{year}{2010}), \bibinfo{pages}{63--77}.
\newblock
\urldef\tempurl%
\url{https://doi.org/10.1177/0018720810366560}
\showDOI{\tempurl}
\newblock
\shownote{PMID: 20653226.}


\bibitem[\protect\citeauthoryear{{Swan}, {Singh}, and {Ellis}}{{Swan}
  et~al\mbox{.}}{2015}]%
        {7164348}
\bibfield{author}{\bibinfo{person}{J.~E. {Swan}}, \bibinfo{person}{G. {Singh}},
  {and} \bibinfo{person}{S.~R. {Ellis}}.} \bibinfo{year}{2015}\natexlab{}.
\newblock \showarticletitle{Matching and Reaching Depth Judgments with Real and
  Augmented Reality Targets}.
\newblock \bibinfo{journal}{\emph{IEEE Transactions on Visualization and
  Computer Graphics}} \bibinfo{volume}{21}, \bibinfo{number}{11}
  (\bibinfo{date}{Nov} \bibinfo{year}{2015}), \bibinfo{pages}{1289--1298}.
\newblock
\showISSN{2160-9306}
\urldef\tempurl%
\url{https://doi.org/10.1109/TVCG.2015.2459895}
\showDOI{\tempurl}


\bibitem[\protect\citeauthoryear{Teather and Stuerzlinger}{Teather and
  Stuerzlinger}{2007}]%
        {10.1145/1328202.1328214}
\bibfield{author}{\bibinfo{person}{Robert~J. Teather} {and}
  \bibinfo{person}{Wolfgang Stuerzlinger}.} \bibinfo{year}{2007}\natexlab{}.
\newblock \showarticletitle{Guidelines for 3D Positioning Techniques}. In
  \bibinfo{booktitle}{\emph{Proceedings of the 2007 Conference on Future Play}}
  (Toronto, Canada) \emph{(\bibinfo{series}{Future Play ’07})}.
  \bibinfo{publisher}{Association for Computing Machinery},
  \bibinfo{address}{New York, NY, USA}, \bibinfo{pages}{61–68}.
\newblock
\showISBNx{9781595939432}
\urldef\tempurl%
\url{https://doi.org/10.1145/1328202.1328214}
\showDOI{\tempurl}


\bibitem[\protect\citeauthoryear{Teather and Stuerzlinger}{Teather and
  Stuerzlinger}{2011}]%
        {10.5555/2013881.2014216}
\bibfield{author}{\bibinfo{person}{Robert~J. Teather} {and}
  \bibinfo{person}{Wolfgang Stuerzlinger}.} \bibinfo{year}{2011}\natexlab{}.
\newblock \showarticletitle{Pointing at 3D Targets in a Stereo Head-Tracked
  Virtual Environment}. In \bibinfo{booktitle}{\emph{Proceedings of the 2011
  IEEE Symposium on 3D User Interfaces}} \emph{(\bibinfo{series}{3DUI ’11})}.
  \bibinfo{publisher}{IEEE Computer Society}, \bibinfo{address}{USA},
  \bibinfo{pages}{87–94}.
\newblock
\showISBNx{9781457700637}


\bibitem[\protect\citeauthoryear{Thumser, Slifkin, Beckler, and
  Marasco}{Thumser et~al\mbox{.}}{2018}]%
        {10.3389/fpsyg.2018.00560}
\bibfield{author}{\bibinfo{person}{Zachary~C. Thumser},
  \bibinfo{person}{Andrew~B. Slifkin}, \bibinfo{person}{Dylan~T. Beckler},
  {and} \bibinfo{person}{Paul~D. Marasco}.} \bibinfo{year}{2018}\natexlab{}.
\newblock \showarticletitle{Fitts’ Law in the Control of Isometric Grip Force
  With Naturalistic Targets}.
\newblock \bibinfo{journal}{\emph{Frontiers in Psychology}}
  \bibinfo{volume}{9} (\bibinfo{year}{2018}), \bibinfo{pages}{560}.
\newblock
\showISSN{1664-1078}
\urldef\tempurl%
\url{https://doi.org/10.3389/fpsyg.2018.00560}
\showDOI{\tempurl}


\bibitem[\protect\citeauthoryear{{Triantafyllidis}, {Mcgreavy}, {Gu}, and
  {Li}}{{Triantafyllidis} et~al\mbox{.}}{2020}]%
        {9076603}
\bibfield{author}{\bibinfo{person}{E. {Triantafyllidis}}, \bibinfo{person}{C.
  {Mcgreavy}}, \bibinfo{person}{J. {Gu}}, {and} \bibinfo{person}{Z. {Li}}.}
  \bibinfo{year}{2020}\natexlab{}.
\newblock \showarticletitle{Study of Multimodal Interfaces and the Improvements
  on Teleoperation}.
\newblock \bibinfo{journal}{\emph{IEEE Access}}  \bibinfo{volume}{8}
  (\bibinfo{year}{2020}), \bibinfo{pages}{78213--78227}.
\newblock
\urldef\tempurl%
\url{https://doi.org/10.1109/ACCESS.2020.2990080}
\showDOI{\tempurl}


\bibitem[\protect\citeauthoryear{Triantafyllidis, Yang, McGreavy, Hu, and
  Li}{Triantafyllidis et~al\mbox{.}}{2020}]%
        {triantafyllidisrobot}
\bibfield{author}{\bibinfo{person}{Eleftherios Triantafyllidis},
  \bibinfo{person}{Chuanyu Yang}, \bibinfo{person}{Christopher McGreavy},
  \bibinfo{person}{Wenbin Hu}, {and} \bibinfo{person}{Zhibin Li}.}
  \bibinfo{year}{2020}\natexlab{}.
\newblock \bibinfo{booktitle}{\emph{Robot intelligence for real-world
  applications}}.
\newblock \bibinfo{publisher}{IET Computing and Networks},
  \bibinfo{address}{UK}. 63 pages.
\newblock


\bibitem[\protect\citeauthoryear{Turk}{Turk}{2014}]%
        {TURK2014189}
\bibfield{author}{\bibinfo{person}{Matthew Turk}.}
  \bibinfo{year}{2014}\natexlab{}.
\newblock \showarticletitle{Multimodal interaction: A review}.
\newblock \bibinfo{journal}{\emph{Pattern Recognition Letters}}
  \bibinfo{volume}{36} (\bibinfo{year}{2014}), \bibinfo{pages}{189 -- 195}.
\newblock
\showISSN{0167-8655}
\urldef\tempurl%
\url{https://doi.org/10.1016/j.patrec.2013.07.003}
\showDOI{\tempurl}


\bibitem[\protect\citeauthoryear{Vaughan, Barany, Sali, Jax, and
  Rosenbaum}{Vaughan et~al\mbox{.}}{2010}]%
        {vaughan2010extending}
\bibfield{author}{\bibinfo{person}{Jonathan Vaughan},
  \bibinfo{person}{Deborah~A Barany}, \bibinfo{person}{Anthony~W Sali},
  \bibinfo{person}{Steven~A Jax}, {and} \bibinfo{person}{David~A Rosenbaum}.}
  \bibinfo{year}{2010}\natexlab{}.
\newblock \showarticletitle{Extending Fitts’ Law to three-dimensional
  obstacle-avoidance movements: support for the posture-based motion planning
  model}.
\newblock \bibinfo{journal}{\emph{Experimental brain research}}
  \bibinfo{volume}{207}, \bibinfo{number}{1-2} (\bibinfo{year}{2010}),
  \bibinfo{pages}{133--138}.
\newblock


\bibitem[\protect\citeauthoryear{Vertegaal}{Vertegaal}{2008}]%
        {10.1145/1452392.1452443}
\bibfield{author}{\bibinfo{person}{Roel Vertegaal}.}
  \bibinfo{year}{2008}\natexlab{}.
\newblock \showarticletitle{A Fitts Law Comparison of Eye Tracking and Manual
  Input in the Selection of Visual Targets}. In
  \bibinfo{booktitle}{\emph{Proceedings of the 10th International Conference on
  Multimodal Interfaces}} (Chania, Crete, Greece) \emph{(\bibinfo{series}{ICMI
  '08})}. \bibinfo{publisher}{Association for Computing Machinery},
  \bibinfo{address}{New York, NY, USA}, \bibinfo{pages}{241–248}.
\newblock
\showISBNx{9781605581989}
\urldef\tempurl%
\url{https://doi.org/10.1145/1452392.1452443}
\showDOI{\tempurl}


\bibitem[\protect\citeauthoryear{Wang and MacKenzie}{Wang and
  MacKenzie}{1999}]%
        {10.1145/302979.302989}
\bibfield{author}{\bibinfo{person}{Yanqing Wang} {and}
  \bibinfo{person}{Christine~L. MacKenzie}.} \bibinfo{year}{1999}\natexlab{}.
\newblock \showarticletitle{Object Manipulation in Virtual Environments:
  Relative Size Matters}. In \bibinfo{booktitle}{\emph{Proceedings of the
  SIGCHI Conference on Human Factors in Computing Systems}} (Pittsburgh,
  Pennsylvania, USA) \emph{(\bibinfo{series}{CHI ’99})}.
  \bibinfo{publisher}{Association for Computing Machinery},
  \bibinfo{address}{New York, NY, USA}, \bibinfo{pages}{48–55}.
\newblock
\showISBNx{0201485591}
\urldef\tempurl%
\url{https://doi.org/10.1145/302979.302989}
\showDOI{\tempurl}


\bibitem[\protect\citeauthoryear{Wang, MacKenzie, Summers, and Booth}{Wang
  et~al\mbox{.}}{1998}]%
        {10.1145/274644.274688}
\bibfield{author}{\bibinfo{person}{Yanqing Wang}, \bibinfo{person}{Christine~L.
  MacKenzie}, \bibinfo{person}{Valerie~A. Summers}, {and}
  \bibinfo{person}{Kellogg~S. Booth}.} \bibinfo{year}{1998}\natexlab{}.
\newblock \showarticletitle{The Structure of Object Transportation and
  Orientation in Human-Computer Interaction}. In
  \bibinfo{booktitle}{\emph{Proceedings of the SIGCHI Conference on Human
  Factors in Computing Systems}} (Los Angeles, California, USA)
  \emph{(\bibinfo{series}{CHI ’98})}. \bibinfo{publisher}{ACM
  Press/Addison-Wesley Publishing Co.}, \bibinfo{address}{USA},
  \bibinfo{pages}{312–319}.
\newblock
\showISBNx{0201309874}
\urldef\tempurl%
\url{https://doi.org/10.1145/274644.274688}
\showDOI{\tempurl}


\bibitem[\protect\citeauthoryear{Ware and Balakrishnan}{Ware and
  Balakrishnan}{1994}]%
        {10.1145/198425.198426}
\bibfield{author}{\bibinfo{person}{Colin Ware} {and} \bibinfo{person}{Ravin
  Balakrishnan}.} \bibinfo{year}{1994}\natexlab{}.
\newblock \showarticletitle{Reaching for Objects in VR Displays: Lag and Frame
  Rate}.
\newblock \bibinfo{journal}{\emph{ACM Trans. Comput.-Hum. Interact.}}
  \bibinfo{volume}{1}, \bibinfo{number}{4} (\bibinfo{date}{Dec.}
  \bibinfo{year}{1994}), \bibinfo{pages}{331–356}.
\newblock
\showISSN{1073-0516}
\urldef\tempurl%
\url{https://doi.org/10.1145/198425.198426}
\showDOI{\tempurl}


\bibitem[\protect\citeauthoryear{Welford}{Welford}{1968}]%
        {welford1968fundamentals}
\bibfield{author}{\bibinfo{person}{A.~T. (Alan~Traviss) Welford}.}
  \bibinfo{year}{1968}\natexlab{}.
\newblock \bibinfo{booktitle}{\emph{Fundamentals of skill}}.
\newblock \bibinfo{publisher}{Methuen}, \bibinfo{address}{London}.
\newblock
\showISBNx{0416030009}


\bibitem[\protect\citeauthoryear{Witmer and Kline}{Witmer and Kline}{1998}]%
        {Witmer:1998:JPT:1246749.1246755}
\bibfield{author}{\bibinfo{person}{Bob~G. Witmer} {and}
  \bibinfo{person}{Paul~B. Kline}.} \bibinfo{year}{1998}\natexlab{}.
\newblock \showarticletitle{Judging Perceived and Traversed Distance in Virtual
  Environments}.
\newblock \bibinfo{journal}{\emph{Presence: Teleoper. Virtual Environ.}}
  \bibinfo{volume}{7}, \bibinfo{number}{2} (\bibinfo{date}{April}
  \bibinfo{year}{1998}), \bibinfo{pages}{144--167}.
\newblock
\showISSN{1054-7460}
\urldef\tempurl%
\url{https://doi.org/10.1162/105474698565640}
\showDOI{\tempurl}


\bibitem[\protect\citeauthoryear{Wobbrock, Cutrell, Harada, and
  MacKenzie}{Wobbrock et~al\mbox{.}}{2008}]%
        {10.1145/1357054.1357306}
\bibfield{author}{\bibinfo{person}{Jacob~O. Wobbrock}, \bibinfo{person}{Edward
  Cutrell}, \bibinfo{person}{Susumu Harada}, {and} \bibinfo{person}{I.~Scott
  MacKenzie}.} \bibinfo{year}{2008}\natexlab{}.
\newblock \showarticletitle{An Error Model for Pointing Based on Fitts’ Law}.
  In \bibinfo{booktitle}{\emph{Proceedings of the SIGCHI Conference on Human
  Factors in Computing Systems}} (Florence, Italy) \emph{(\bibinfo{series}{CHI
  ’08})}. \bibinfo{publisher}{Association for Computing Machinery},
  \bibinfo{address}{New York, NY, USA}, \bibinfo{pages}{1613–1622}.
\newblock
\showISBNx{9781605580111}
\urldef\tempurl%
\url{https://doi.org/10.1145/1357054.1357306}
\showDOI{\tempurl}


\bibitem[\protect\citeauthoryear{Wobbrock, Jansen, and Shinohara}{Wobbrock
  et~al\mbox{.}}{2011}]%
        {10.1145/1978942.1979183}
\bibfield{author}{\bibinfo{person}{Jacob~O. Wobbrock}, \bibinfo{person}{Alex
  Jansen}, {and} \bibinfo{person}{Kristen Shinohara}.}
  \bibinfo{year}{2011}\natexlab{}.
\newblock \showarticletitle{Modeling and Predicting Pointing Errors in Two
  Dimensions}. In \bibinfo{booktitle}{\emph{Proceedings of the SIGCHI
  Conference on Human Factors in Computing Systems}} (Vancouver, BC, Canada)
  \emph{(\bibinfo{series}{CHI ’11})}. \bibinfo{publisher}{Association for
  Computing Machinery}, \bibinfo{address}{New York, NY, USA},
  \bibinfo{pages}{1653–1656}.
\newblock
\showISBNx{9781450302289}
\urldef\tempurl%
\url{https://doi.org/10.1145/1978942.1979183}
\showDOI{\tempurl}


\bibitem[\protect\citeauthoryear{Zaroff, Knutelska, and Frumkes}{Zaroff
  et~al\mbox{.}}{2003}]%
        {10.1167/iovs.02-0361}
\bibfield{author}{\bibinfo{person}{Charles~M. Zaroff}, \bibinfo{person}{Magosha
  Knutelska}, {and} \bibinfo{person}{Thomas~E. Frumkes}.}
  \bibinfo{year}{2003}\natexlab{}.
\newblock \showarticletitle{{Variation in Stereoacuity: Normative Description,
  Fixation Disparity, and the Roles of Aging and Gender}}.
\newblock \bibinfo{journal}{\emph{Investigative Ophthalmology \& Visual
  Science}} \bibinfo{volume}{44}, \bibinfo{number}{2} (\bibinfo{date}{02}
  \bibinfo{year}{2003}), \bibinfo{pages}{891--900}.
\newblock
\showISSN{1552-5783}
\urldef\tempurl%
\url{https://doi.org/10.1167/iovs.02-0361}
\showDOI{\tempurl}


\bibitem[\protect\citeauthoryear{Zeng, Hedge, and Guimbretiere}{Zeng
  et~al\mbox{.}}{2012}]%
        {doi:10.1177/1071181312561207}
\bibfield{author}{\bibinfo{person}{Xiaolu Zeng}, \bibinfo{person}{Alan Hedge},
  {and} \bibinfo{person}{Francois Guimbretiere}.}
  \bibinfo{year}{2012}\natexlab{}.
\newblock \showarticletitle{Fitts’ Law in 3D Space with Coordinated Hand
  Movements}.
\newblock \bibinfo{journal}{\emph{Proceedings of the Human Factors and
  Ergonomics Society Annual Meeting}} \bibinfo{volume}{56}, \bibinfo{number}{1}
  (\bibinfo{year}{2012}), \bibinfo{pages}{990--994}.
\newblock
\urldef\tempurl%
\url{https://doi.org/10.1177/1071181312561207}
\showDOI{\tempurl}


\bibitem[\protect\citeauthoryear{Zhai, Milgram, and Buxton}{Zhai
  et~al\mbox{.}}{1996}]%
        {10.1145/238386.238534}
\bibfield{author}{\bibinfo{person}{Shumin Zhai}, \bibinfo{person}{Paul
  Milgram}, {and} \bibinfo{person}{William Buxton}.}
  \bibinfo{year}{1996}\natexlab{}.
\newblock \showarticletitle{The Influence of Muscle Groups on Performance of
  Multiple Degree-of-Freedom Input}. In \bibinfo{booktitle}{\emph{Proceedings
  of the SIGCHI Conference on Human Factors in Computing Systems}} (Vancouver,
  British Columbia, Canada) \emph{(\bibinfo{series}{CHI '96})}.
  \bibinfo{publisher}{Association for Computing Machinery},
  \bibinfo{address}{New York, NY, USA}, \bibinfo{pages}{308–315}.
\newblock
\showISBNx{0897917774}
\urldef\tempurl%
\url{https://doi.org/10.1145/238386.238534}
\showDOI{\tempurl}


\bibitem[\protect\citeauthoryear{{Zheng}}{{Zheng}}{2013}]%
        {6335488}
\bibfield{author}{\bibinfo{person}{Y. {Zheng}}.}
  \bibinfo{year}{2013}\natexlab{}.
\newblock \showarticletitle{An Efficient Algorithm for a Grasp Quality
  Measure}.
\newblock \bibinfo{journal}{\emph{IEEE Transactions on Robotics}}
  \bibinfo{volume}{29}, \bibinfo{number}{2} (\bibinfo{year}{2013}),
  \bibinfo{pages}{579--585}.
\newblock


\end{thebibliography}
\end{document}